# GENERAL LAW OF GROWTH AND REPLICATION, GROWTH EQUATION AND ITS APPLICATIONS

YURI K. SHESTOPALOFF

SegmentSoft Research Lab, 61 Danby Ave, Toronto, Ontario, M3H 2J4 Canada

shes169@yahoo.ca

**Abstract.** We present significantly advanced studies of the previously introduced physical growth mechanism and unite it with biochemical growth factors. Obtained results allowed formulating the general growth law which governs growth and evolutional development of all living organisms, their organs and systems. It was discovered that the growth cycle is predefined by the distribution of nutritional resources between maintenance needs and biomass production. This distribution is quantitatively defined by the growth ratio parameter, which depends on the geometry of an organism, phase of growth and, indirectly, organism's biochemical machinery. The amount of produced biomass, in turn, defines the composition of biochemical reactions. Changing amount of nutrients diverted to biomass production is what forces organisms to proceed through the whole growth and replication cycle. The growth law can be formulated as follows: the rate of growth is proportional to influx of nutrients and growth ratio. Considering specific biochemical components of different organisms, we find influxes of required nutrients and substitute them into the growth equation; then, we compute growth curves for *amoeba*, *wild type fission yeast*, *fission yeast's* mutant. In all cases, predicted growth curves correspond very well to experimental data. Obtained results prove validity and fundamental scientific value of the discovery.

## Contents

**1. Introduction**

The idea of existence of a general growth law that governs the growth of all living organisms, and its mathematical representation, the growth equation, has been previously introduced in several articles and books[1,2,3]. However, the previous publications left many issues open. In particular, using geometrical considerations, the author proved that the specific influx of nutrients generally increases during growth, but no reliable connection of nutrient influx to the biochemistry of organisms was established. Another important issue was interpretation of some terms of the growth equation, in particular the growth ratio, which turned out to be of a fundamental importance as subsequent study showed. Once these issues were resolved, we were able to formulate the general growth and replication law, which states that "*the growth cycle of a living organism and appropriate changes in composition of biochemical reactions are defined by the distribution of nutrients between maintenance needs and biomass production in such a way that the fraction of nutrients directed to biomass production at any given moment is equal to the growth ratio (which is a monotonically decreasing function), so that the growth rate is proportional to influx of nutrients and the growth ratio.*"

    The heuristic nature and importance of fundamental laws requires their thorough verification, which is the main subject of this article. Another theme relates to practical applications of the discovered general growth mechanism. History shows that eventually humankind reaps huge benefits from such discoveries. In this paper, we consider a few applications of introduced concepts to long standing growth and replication problems. Other examples will follow in separate articles. Since the general growth law is an influential factor for any living organism, it can be applied to a wide variety of pertinent practical problems.

    It is important that from the very beginning the reader could understand that we do not consider some mathematical data fit model, but a *fundamental law* of nature. Although we use some experimental data for *generalizations*, such as for finding value of a spare growth

capacity, the actual growth curves are calculated based *entirely* on the growth equation, whose input parameters are *completely* defined by biochemical properties of organisms and their geometry. Only *then* we compare the computed growth curves to experimental data. The value of the time scaling coefficient, which is the only adjustable parameter, does not change the form of the growth curve but scales it along the abscissa (time axis).

The aforementioned growth ratio parameter plays important role in the general law of growth. However, the growth ratio is not a magic number, but an objective parameter, a mathematical expression of interaction and optimal balance of three components:
1. Nutrient supply abilities (they depend on nutrients availability in the environment and ability of organisms to acquire these nutrients).
2. Nutritional requirements of organism's mass (of course, associated with volume and geometry), since organism has to maintain functioning of existing mass and produce biomass using supplied nutrients.
3. Specifics of biochemical machinery of an organism.

Numerically, the value of the growth ratio defines an optimal use of nutritional resources when both maintenance of existing biomass and synthesis of a new one are adjusted at highest possible capacities. The fraction of nutrients that is used for biomass production at any given moment is equal to the growth ratio. It is quite logical that such an optimization mechanism had to be developed during evolution. If too many resources go to maintenance, then organisms would grow slowly, thus losing their competitive advantage. If too much nutrients are used for biomass synthesis, then the organism's biochemical machinery will be under stress, since not much nutrients will be left for maintenance. Consequently, all functions of an organism, including biomass production, will be impaired and the growth as a whole will not be optimal.

For those who know classical mechanics, a good analogy of the growth ratio can be a Hamiltonian, which defines the motion of a system of physical objects given certain initial conditions. In normal circumstances, there is only one unique way how all these objects will move, although at a first glance they can go along any trajectory independently of each other or at least that there are several scenarios of collective motion. This certainty is due to collective action of physical *laws* and interdependence of objects through acting forces and system's characteristics. In the case of growth of living organisms, we *do have* similar certainty due to action of physical, biochemical and other laws. A minor problem is that we have to know these laws. The discovered general growth law is such a fundamental law of nature that governs the growth of all living species and its components, such as organelles, organs and systems, at all levels. Fig. 1 shows the action of the general growth mechanism.

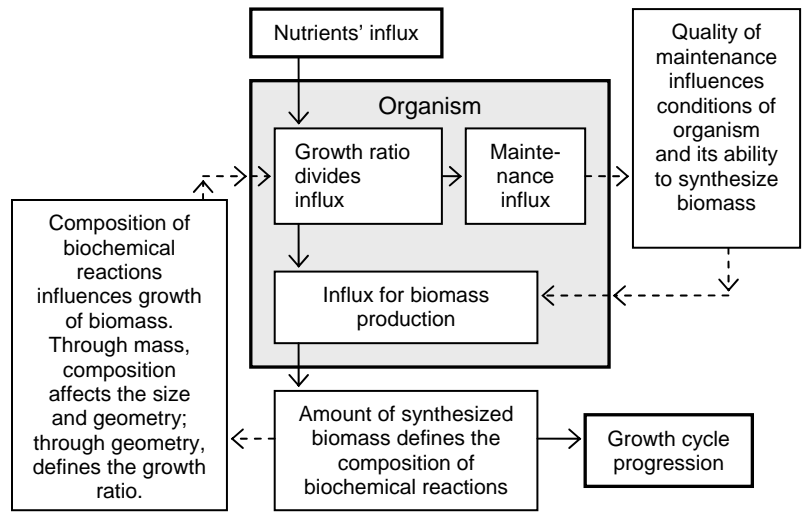

Fig. 1. Growth cycle regulation and progression.

## 2. Biochemical machinery

Biochemical mechanisms of growth and replication, first of all DNA, are viewed by many as the primary mechanisms that organize and direct growth and support organic life in general. Of course, biochemical mechanisms are important. However, they are not the only growth factors, but interact and work in close cooperation with physical, electromagnetic and other mechanisms, under the overall guidance of the general growth mechanism. Certainly, without biochemical reactions no living organism would ever appear. However, without general growth mechanism, or growth law, no evolutionary development and mere existence of living organisms would be possible either.

In order to understand the hierarchy of growth mechanisms one can use a simple criterion - what mechanism is more general? In Ref. 4, the authors say that "*the underlying "cell cycle engine" is remarkably conserved*", despite a different set of particular biochemical reactions in different cells. Numerous studies in molecular biology convincingly prove this thesis. So, maybe the major biochemical building blocks are more general than the general growth law? Not quite so. As we know, even the omnipresent DNA can be substituted in some microorganisms, while the general growth law is equally applicable to all living species and effectively governs its evolutionary development[5] *and* the growth and replication of individual organisms. (Certainly, it does this together with biochemical mechanisms.)

A formation of biochemical machinery by and large is a *response* to conditions imposed by environment. One of the most important tasks that every organism has to accomplish relates to optimal distribution of resources for supporting existing structure and, at the same time, providing its growth. This distribution is not an arbitrary process. The general growth law, in this regard, is a particular realization of this fundamental distribution mechanism applied to living organisms.

Let us consider an evolutional development of the world we know. Atoms and molecules permanently combine into more complex structures and sooner or later disintegrate. The number of possible permutations even for a small number of basic elements is *enormous*, as well as the speed of creation of such combinations. Among these myriads of combinations, more stable structures occur, which through selection can gradually become more numerous. It is important to first form a simple reproduction mechanism in order to begin improving it, as opposed to creating a complex structure every time from scratch. Was it possible through random combinations of different atoms and molecules? The answer is obvious, because this is what we see around today. However, the problem is *conceiving* this event, which might be an obstacle unless one understands how enormous the amount of permutations that *really* happened on a molecular level is, and how many active forces working collectively and very effectively participate in formation of molecular structures. Knowing these factors, we can calculate what time is required in order to obtain such a reproductive unit with high probability (since we know the structure of basic reproduction blocks, major mechanisms forming the biochemical structures, and we can make credible assumptions about the environmental conditions). Then, we can compare the result to the answer that we know – of the order of billion years. Intuitively, given the complexity of the task and knowing the nature of atomic forces, it is reasonable to assume that the construction of the first reproduction block should take the longest time. Once nature obtained these reproduction building blocks, it began manipulations with *blocks*, not separate atoms. The number of components drastically decreased, by several orders of magnitude, and so the development of organic life accelerated. The higher a form of life is, the less high level building blocks we have and less manipulations with major building blocks are allowed, since more complex systems are more sensitive to disturbances and divergences, because their functioning depends on many different factors working together. The more complex a system is, the more it is vulnerable, so that only a small fraction of possible combinations of major building blocks (let us say, on the system and organ levels) represents sustainable organisms which can adequately function and compete, while in case of simple molecular compositions the spectrum of their properties is continuous and more uniform, because of the abundance and sustainability of many simple chemical compositions.

Let us elaborate on the appearance of first reproductive units, since this is what might puzzle many people. Indeed, origination of reproductive units is a *qualitative* jump in the evolutionary process. However, it is not as unlikely as many might be inclined to think. Once certain stable temperature conditions, which are very likely to occur in water, and certain mineral composition that is needed for synthesis of organic molecules (a few basic elements) are created, electromagnetic forces begin combining these elements and smaller stable organic compounds into more complex molecular structures. Once a complex molecule is created, in certain, likely to occur on the Earth, conditions it can work as a template for creation of another similar molecule. Those who know basics of biochemistry would immediately recognize this arrangement as the base mechanism of biochemical synthesis of cell proteins and other components.

Two conditions must be fulfilled in order for such a process to become possible. First, the template molecule has to expose sites to which smaller organic molecules can bind. Second, bonds between the components of a copied complex molecule have to be *stronger* than the bonds between the template and copied molecule, or a chemical agent that is targeting specific bonds between the template and copied molecules does not influence the bonds of template and created molecules. At a glance, within the scope of organic chemistry, fulfillment of these conditions is quite possible, even if we would be unaware of such existing mechanisms.

We can ask, why similar duplication mechanism was not developed for simple non-organic molecules, like cuprum dioxide? The answer is this. Those molecules are held by *strong* electromagnetic forces, lets us say, forces of the first level. Complex organic molecules, in this regard, are held by *weaker* electromagnetic forces, of the second level, which originate as the result of lack or excess of electrons on the *outer* boundaries of less complex organic molecules and / or as molecular bonds (for instance, dipoles). Bonds between the original complex molecule and its copy represent even weaker electromagnetic forces of the third level. In order such a copy of non-organic molecule could be created and later disconnected from the template molecule, the bonds have to be weak enough, which is not the case for primary chemical compositions.

The other side of any growth is inevitable destruction, which comes in different forms. (In fact, the general growth law is applicable to shrinking systems as well, which we will study in this work.) For instance, environmental conditions may deteriorate to such an extent that no organic life becomes possible.

### 3. Introducing the growth equation

In this paper, besides the previously considered growth equation for organisms that obtain nutrients through the surface, we will introduce the growth equation for organisms that receive nutrients in other ways, such as through a fruit's stem, of blood vessels, or lungs. Also, we introduce systems of growth equations and appropriate constraints for multicellular organisms, as well as for complex organisms such as humans that have many organs and systems.

In simple growth scenarios, let us say, for a unicellular microorganism that receives nutrients through its membrane, the growth equation has the following form.

$$p_c(X)dV(X,t) = k(X,t) \times S(X) \times \left(\frac{R_S}{R_V} - 1\right)dt \qquad (1)$$

Here, $X$ is a vector that represents coordinates of some point in a cell; $p_c$ is the density of the cell's substance measured in $kg/m^3$ that can generally depend on the coordinate vector $X$; $t$ is time; $k$ is the specific influx (amount of nutrients per unit of surface per unit of time, so that it is measured in $kg/(m^2 \times \text{sec})$); $p_c(X)dV(X,t)$ is the change in the cell's mass; $S(X)$ is the cell's surface area; $R_S$ and $R_V$ denote the relative surface and volume accordingly, which can also depend on the coordinate vector $X$ and time $t$. In general, the density of the cell $p_c$ can also depend on the coordinate vector $X$.

The growth equation has the following interpretation. The left part represents the mass increment during time interval $dt$. The right part represents the total influx of nutrients $K$ through the surface (the term $K = k(X,t) \times S(X)$, measured in $kg/\sec$), multiplied by the value of the growth ratio $G_R = (R_S / R_V - 1)$, which thus defines the *fraction* of the total influx that is used for biomass synthesis. For multicellular organisms, one should use the *total* influx; for instance, when nutrients come through a fruit stem or blood vessels (we will consider this case in a separate section).

The *growth ratio* depends on the geometrical characteristics of an organism and, indirectly, on the organism's biochemical machinery through the *maximum possible size* that can be achieved for a particular organism. Let us assume that the nutrients' availability and the biochemical arrangement of some organism, which receives nutrients through the surface, allows the organism to grow to a maximum volume of $V_0$ with a corresponding surface area of $S_0$. Then, we can define the *dimensionless* relative surface $R_S$ and the relative volume $R_V$ as follows.

$$R_S = \frac{S(V)}{S(V_0)} \tag{2}$$

$$R_V = \frac{V}{V_0} \tag{3}$$

The growth ratio $G_R$, which is also a dimensionless value, is defined as follows.

$$G_R = \frac{R_S}{R_V} - 1 \tag{4}$$

In Refs. 1-3, the growth ratio was defined as $G_R = R_S / R_V$. However, it is more convenient to use (4). Although the growth ratio is introduced through geometrical characteristics, it is linked to biochemistry of organisms in two ways. First, it defines how much nutrients the biochemistry machinery uses for biomass production and how much for maintenance. Second, it depends on the maximum possible size associated with $V_0$, which depends on the efficiency of biochemical machinery to process nutrients and synthesize biomass.

As an example, let us consider a cell model that has a disk like shape. Using (2) – (4), we can find the relative surface, the relative volume and the growth ratio for the disk as follows.

$$R_S = \frac{r(r+H)}{R(R+H)} \qquad (5)$$

$$R_V = \frac{r^2}{R^2} \qquad (6)$$

$$G_R = \frac{R_S}{R_V} - 1 = \frac{R(r+H)}{r(R+H)} - 1 \qquad (7)$$

where $r$ is the current radius of the disk, $R$ is the radius of the disk corresponding to the maximum possible volume, $H$ is the disk's height.

*The maximum possible size parameter.* The maximum possible size may look like an ambiguous parameter. How can one know the maximum possible size *before* the growth completes? In fact, there are many a priori known constraints that allow finding the maximum possible size with sufficient accuracy for practical purposes. For instance, composition of biochemical reactions is largely defined by the type of an organism, its initial size, nutrients' availability and their chemical composition. Also, there are certain stable relationships between the geometry and the biochemistry of organisms that allow finding the maximum possible size too. For example, it was earlier discovered[1,2,3] that at least two major types of growth mechanisms were evolutionarily developed. One is when organisms use almost the *whole* growth cycle predefined by the growth equation, such as *amoeba* or *S. cerevisiae*. In this case, the ending size is defined by a certain asymptotic value which can generally be found once we know a particular type of an organism, its initial size, nutrients composition and their concentration. Certainly, nutrients concentration and content can change during the growth; biochemical machinery of organisms of the same type can differ, and other factors can interfere including random ones. However, the good news is that this new information can be accommodated and so the maximum possible size can be corrected. The precise value of the maximum possible size will not be known until the growth ends, but we can predict it with reasonable accuracy.

The second type of growth is when organisms use only *the fastest* part of the possible growth curve, such as *S. pombe* or *E. coli*. Mathematically, the point of switching to the division phase for such organisms corresponds to *inflection* point of the growth curve. In order to switch to division earlier, these organisms have more sophisticated biochemical machinery that is also supported by an appropriate geometrical form, such as a cylinder like shape of *S. pombe* or *E. coli*. This is because in the case of elongated forms, the inflection point is better expressed than for organisms that grow in all dimensions, such as a spherical bacterium. For this type of growth, the situation with a preliminary evaluation of the maximum possible size is even better, because, in fact, we have two characteristic points. One is the inflection point, which corresponds to the *actual* size of an organism when it starts division. The other one is the maximum *possible* size, which is related to the actual size through the spare growth capacity. The latter apparently does not vary much, at least for the same type of organisms, if not for the same type of growth scenario (which is very likely given the fact that biochemical growth machinery is very similar across different microorganisms[4].

## 4. Finding influx of nutrients based on biochemical considerations

One of the main parameters of the growth equation (1) is specific influx of nutrients. Analysis of the literature, modeling and experiments done in Ref. 5 discovered two types of compositions of biochemical reactions responsible for protein and RNA synthesis. Note that these types of biochemical compositions do not coincide with two growth scenarios described in the previous section. For instance, *amoeba* and *S. cerevisiae* use the full growth cycle, but the rate of RNA synthesis in *S. cerevisiae* is about twice of the rate of protein synthesis, while in *amoeba* they are about the same.

### 4.1. *Growth rate and mass increase*

Let us denote the growth rate as $\mu$. The rate of synthesis is defined as the number of doublings of cells' mass per hour. (In some instances, the same value can be defined as the time that is required to double the *number* of cells.) The relative growth of some organism's mass *m*, depending on time, can be expressed as an exponent with the base of two.

$$m_1(t)/m_{01} = 2^{t\mu} \tag{8}$$

where *t* is time, $m_{01}$ is the beginning mass at point $t = 0$.

If the growth rate for the increase of some other organism's mass $m_2$ is double the previous one, that is $2\mu$, then we can write:

$$m_2(t)/m_{02} = 2^{t2\mu} \tag{9}$$

$$m_2(t)/m_{02} = (m_1(t)/m_{01})^2 \tag{10}$$

Here, $m_{02}$ is the beginning mass at point $t = 0$.
In other words, the relative increase of mass in the second case, when the growth rate is $2\mu$, is the *square* of the relative increase of mass in the first case when the growth rate is $\mu$.

### 4.2. *Ribosome and protein rates of synthesis*

Ribosomes are of paramount importance for the cell growth and maintenance – ribosomes synthesize proteins. The authors of Ref. 6 summarize, referring to several works such as Ref. 7 that "*Ribosome synthesis dominates a growing cell's economy, accounting for more than 50% of total transcription in budding yeast and mammalian cells*". In Ref. 8 we find, "*the synthesis of ribosome accounts for the cell's single largest expenditure of biosynthetic energy*".

In Ref. 7, the author obtained himself, and cite other works with similar results, that "*rate of synthesis of protein increases exponentially through the cell cycle*"; and also that "*the rate of RNA synthesis increases exponentially through the cell cycle*". Although it was first thought that these rates are close in value, the following works discovered that the rate of ribosome and RNA synthesis can be substantially higher. For instance, the authors of Ref. 9 discovered a *doubling* in the rate of rRNA synthesis and poly(A)-containing RNA in *S. cerevisiae* during S phase, and the preservation of this high rate through the growth cycle. The same results were reported for *Saccharomyces pombe*[9,10]. Since these results were obtained for *synchronically* growing microorganisms, they are applicable to an *individual* cell.

On the other hand, there are many works that consider the growth of *cultures* depending on the growth rate, such as *E. coli* and *Salmonella typhimurium*[11]. The change of the growth rate is achieved by using different nutrient environments. While the rate of protein synthesis, according to authors, is proportional to the growth rate, they acknowledge that "*the ribosome synthesis rate is increasing approximately with the square of the growth rate*". They also note that "*…at rate ranging from 0.2 to 2.4 doublings/hr, shows that the quantity of ribosomal RNA increases more sharply with the growth rate than does total RNA*". They also found that "*the number of ribosomes is proportional to the rate of growth and of protein synthesis*", which means that larger cells contain a greater number of ribosomes per unit of volume, in proportion to the rate of growth. The same results were obtained by authors for another series of experiments when they considered nutritional shifts in the culture's medium: "*rate of protein synthesis goes hand in hand with an increase in the number of mature ribosomes*". This very well agrees with the previously discussed results about the doubling of rate of RNA synthesis during the growth cycle in individual cells. Although the results for a culture cannot be directly applied to an individual cell, the ability of growing cells to increase the ribosome synthesis rate as a square of the cell's growth rate in a culture, in the presence of more nutrients, confirms the existence of an accelerated ribosome synthesis mechanism compared to the rate of protein synthesis.

The second part of the last quote, which refers to the constant efficiency, was later corrected by other authors. We should note that the rate at which ribosomes function actually has some range of flexibility depending on other factors, for instance, nutritional. In Ref. 6, the authors refer to several works, summarizing the result as follows: "*Nutrient sensing pathway controls not only the rate at which ribosomes are produced and the cytoplasmic ribosome concentration, but also the rate at which ribosomes function*".

We have to add that the note about the doubling of synthesis rate of ribosomes and rRNA compared to the protein synthesis rate does not apply to tRNA, whose rate of synthesis is about the same as for protein. In Ref. 11, the authors acknowledge that "*the tRNA/DNA ratio does not change significantly with the growth rate*". They present approximations of experimental dependencies of DNA and RNA mass on the growth rate as follows.

$$\log_{10}(RNA) = 0.45\mu + 1.31$$

$$\log_{10}(DNA) = 0.23\mu + 0.75$$

We can see that exponents presenting the rate of synthesis of RNA and DNA differ by about two times, that is $0.45\mu / 0.23\mu \approx 2.0$. Given the aforementioned close relationship between the masses of tRNA and DNA, this also means that the rates of synthesis of tRNA and RNA also differ by approximately two times.

tRNA constitutes a few percent of the overall cell's mass, while rRNA accounts for a larger part of the overall synthesis[12]: "*Ribosome synthesis is a massive consumer within the economy of the yeast cell, where rRNA transcription represents ~60% of total transcription and RP mRNA transcription represents ~50% of the total Pol II transcription initiation events*" (RP stands for "ribosomal proteins"). Similar results were presented in Ref. 7.

The rate of mRNA synthesis rather mirrors that of rRNA. There are many indicators for such an assumption. In particular, in Ref. 13, the authors acknowledge that "*Translation of r-protein mRNAs during Drosophila development closely parallels rRNA transcription*", and "*The rate of ribosome synthesis in Drosophila ovaries is probably the highest of any tissue at any developmental time*".

Additional proofs of the above inferences can be found in Ref. 14. The authors acknowledge the following:
"*The ribosome synthesis rate increases approximately with the square of the growth rate.*"
"*in moderate to fast growing bacteria, ... rRNA synthesis per unit protein increases with the square of the cellular growth rate, and ribosomes therefore accumulate in proportion to the growth rate*".
"*The proportionality between ribosome concentrations and growth holds only in the medium to fast growing range, but not in the slow growth conditions.*"

The last observation was also mentioned in Ref. 11, in which the readers can see pictures that demonstrate a big difference in concentration of ribosomes in fast and slow growing cells. This fact is important since it explains difference in growth curves for fast and slow growing cells, although we will not elaborate on that in this paper.

Ref. 15 presents the graph (figure 2) that the energy requirements for RNA synthesis increase when the growth rate increases. This also confirms that the rate of synthesis of RNA and protein can be different.

**4.3. *Rate of synthesis and influx***

Let us find influx $K_p$, which is required for protein synthesis, depending on the rate of increase of protein mass. Differentiating (8) with respect to time, we find:

$$K_p = \frac{C_1}{m_{01}} \times \frac{dm_1(t)}{dt} = C_1 \mu 2^{t\mu} \ln 2 = C_1 \mu \ln 2 \frac{m_1(t)}{m_{01}} \qquad (11)$$

where $C_1$ is a constant coefficient, such that the dimension of $K$ is $[kg/\sec]$.

Thus, the influx required for protein synthesis is proportional to the cell's mass. If the cell density is constant, then we can also say that influx is proportional to the cell's volume.

As we can see from the above discussion, for some organisms, the double value of rate of RNA synthesis, compared to the rate of protein synthesis, is a reasonable assumption. However, we should not rule out the possibility that, for other organisms, it could be less, given some published results.

If we assume that the rate of RNA synthesis is double of the rate of protein synthesis, then, taking into account (12) and (13), we can find the influx $K_R$ that is required for the RNA synthesis as follows.

$$K_R = \frac{C_2}{m_{02}} \times \frac{dm_2(t)}{dt} = C_2 2\mu 2^{t2\mu} \ln 2 = C_2 2\mu \ln 2 \frac{m_2(t)}{m_{02}} = C_2 2\mu \ln 2 \left(\frac{m_1(t)}{m_{01}}\right)^2 \quad (12)$$

where $C_2$ is also a constant coefficient. So, the influx required for RNA synthesis is proportional to the *square* of protein mass.

Now, we have to find the dependence of the total influx into the cell through the membrane on the cell's mass. Note that there is no direct relationship between the consumed nutrients for the synthesis of some cell's components and the relative content of these components within the cell. The relative RNA content may remain the same throughout the cell cycle, which is usually the case, although the rate of RNA synthesis may be higher than that of protein. This happens because of the different turnover rates for different constituents of the cell. If the turnover rate for RNA is higher than the rate of protein synthesis, than the relative content of RNA and protein components may remain the same throughout the cell cycle despite their different rates of synthesis.

However, the important thing is that the rate of synthesis is directly related to the amount of nutrients required for the synthesis of particular components; in other words, it is more linked to nutrients' influx. As the first approximation, we may use the law of conservation of mass and assume that the total influx that is used for protein synthesis (and for other substances synthesized at about the same rate as protein) is proportional to the cell mass. This is the approach which is used in metabolic flux analysis[16]. On the other hand, the cell does many other things besides the synthesis of protein, such as proofreading of DNA and protein, proton leakage across membrane, etc.[15]. All these numerous activities require energy and consequently nutrients.

The influx of nutrients that is required for protein synthesis is, in turn, proportional to the cell mass, because protein constitutes a relatively stable and also the largest part, (about 55%) of the total cell mass. Accordingly, the influx of nutrients for synthesis of ribosomes, based on (12), is proportional to the square of mass. The efficiency of biochemical machinery is about the same for synthesis of proteins and RNA, because both processes use the transcription and translation mechanisms of the same nature. So, we can assume that the influx of nutrients that is required for synthesis at the beginning is proportional to relative contents of protein and RNA. As we mentioned already, the relative content of different components may remain the same through the whole growth cycle, although rates of synthesis may differ. This is the case for protein and RNA content in many organisms, in which the component that is synthesized with a higher rate decays faster. Using these considerations, we can define the specific influx required for protein and RNA synthesis as follows.

$$k(m_C) = \frac{K(m_C)}{S(m_C)} = \frac{C_k \left( \frac{m_{P0}}{m_{C0}} \left( \frac{m_C}{m_{C0}} \right) + \frac{m_{R0}}{m_{C0}} \left( \frac{m_C}{m_{C0}} \right)^2 \right)}{S(m_C)} \tag{13}$$

Here, $m_C$ is the current mass of the cell; $m_{C0}$ is the cell's mass at the beginning; $m_{P0}/m_{C0}$ is the fraction of mass of protein and other cell's components whose rate of synthesis is proportional to the relative cell mass $m_C/m_{C0}$; $m_{R0}/m_{C0}$ is a similar fraction of RNA, whose rate of synthesis is proportional to the square of relative cell mass; $C_k$ is a constant coefficient that has dimension $[kg/\sec]$.

Note that $m_{P0}$ and $m_{R0}$ in general case may vary depending on the phase of the growth cycle, although below we assume them to be constant based on the results of experimental studies, such as the ones presented in Ref. 11.

We can also view (13) from the perspective of distribution of nutrients that, on one hand, are required to increase and support a *production* facility, which in our case is the ribosome machinery that produces proteins. On the other hand, we need "raw materials" for this production facility in order for our plant to produce the product it was designed to manufacture. In our case, these are nutrients for protein synthesis.

### 4.4. *Nutrient influx required to support transportation and signaling networks*

Nutrients have to be delivered to the site of synthesis. Also, we should not forget about energy and synthesis of substances required for signaling networks, since they constitute an inherent part of any living organism. When the rates of RNA and protein synthesis are the same, then the first term in (13) is equally applicable to both substances. However, this will be the minimum nutrient influx requirement, since it only provides nutrients for synthesis of protein and RNA, but does not take into account additional energy and nutrient requirements. So, we can write for the minimum total influx (amount of nutrients per unit of time).

$$K_{\min} = (C_p v + C_r v) \tag{14}$$

Here, $C_r$ and $C_p$ are the weighting coefficients (units of measure $[kg/\sec]$) corresponding to fractions of influx that are used to synthesize protein and RNA. Note that we use the *dimensionless* volume $v$ in (14), which is the ratio of the current volume to the initial volume, so that $v$ is a dimensionless value greater than one.

When the rate of RNA synthesis is twice the rate of protein synthesis, we have

$$K_{\min}(v) = \left(C_p v + C_r v^2\right) \tag{15}$$

It is quite reasonable to assume that transportation and signaling costs are proportional to the distance the nutrients and the waste (which is associated with nutrients) have to be transported and the signals transmitted (we will obtain indirect experimental proof of the validity of this assumption later). We analyze the linear case first, such as the growth of *S. pombe*, which has a cylinder like shape and grows in one dimension, that is, it only elongates. In this case, the signaling and communication pathways elongate as the cell's length increases. In order to better understand our considerations, we may use a railroad construction analogy. Suppose some company builds a long railway, supplying all required resources, such as materials, workers, machinery, food for personnel, etc. from the starting location. We assume that the weight of materials that is required to build one unit of railway length is $w$. The price of material, labor and construction costs per unit of weight is $p_m$. The price of delivering one unit of materials to a distance of one unit is $p_t$, so that the price of delivery is proportional to the traveled distance $l$, which is a reasonable assumption. Then, the total price $dP$ of building a new stretch of railway $dl$ at distance $l$ from the starting point is defined as follows.

$$dP = (wp_m + p_t wl)dl \tag{16}$$

Solving this differential equation with respect to $P$, we obtain the following.

$$P = p_m wl + (1/2) p_t wl^2 \tag{17}$$

We can see from (17) that the transportation costs are proportional to the *square* of the railroad length.

When a cell like *E. coli* grows lengthwise, the volume is proportional to the relative increase in length, so that (15) can be rewritten for the minimum total influx as follows.

$$K_{\min}(L) = \left(C_p L + C_r L^2\right) \tag{18}$$

Similarly to (12) and (13), we assume that $L$ in (14) is a *dimensionless* length, the ratio of the current length to the initial length. Now let us add to (18) the "infrastructure" expenditures analogous to (16) and (17). In other words, we have to add the transportation costs to every unit of mass of influx that is required for protein and ribosome synthesis. Similarly to the second term in (16), we can write for the total influx that includes transportation costs the following.

$$dK(L) = K_{\min}(L)dL \tag{19}$$

which is a mathematical expression of the fact that the amount of additional nutrients that are required to transport one unit of influx into the destination point of synthesis is proportional to the travel distance. Substituting (18) into (19) and solving it, we obtain the following.

$$K(L) = \left(C_p L^2 + C_r L^3\right) \qquad (20)$$

Similarly, we can consider two- and three-dimensional growth. For instance, for the disk, whose height remains constant during growth, and the rates of protein and RNA synthesis are the same, we find:

$$K = \frac{C}{H}\left(\frac{v^{3/2}}{\sqrt{H}}\right) \qquad (21)$$

where $v = V/V_b$, and $V_b$ is the beginning disk volume; $C$ is a constant coefficient. Note that without the infrastructure costs influx is proportional to increase of relative volume.

For a sphere we have $K = Cv^{4/3}$, when the rates of protein and ribosome synthesis are the same. If the rate of RNA synthesis is double that of protein, then

$$K(V) = \left(C_p (V/V_b)^{4/3} + C_r (V/V_b)^{7/3}\right) \qquad (22)$$

Generalizing these considerations, we can write the growth equation that takes into account the infrastructure "toll" for all cell components as follows.

$$p_c(X)dV(X) = k_{\min}(X) \times \left(r(X)/r_0(X)\right) S(X) \times \left(\frac{R_S}{R_V} - 1\right) dt \qquad (23)$$

where the new variable $r$ is the distance that the synthesized and raw substances have to be transported; $r_0$ is the initial transport distance in the same direction; $k_{\min}$ is the minimum required nutrients' influx without "infrastructure costs". Given the fact that much of the activity in a cell is directed from the periphery to the center (nuclei) and vice versa, in many instances it will be reasonable to assume that $r$ in (23) is the distance from the center to the elementary volume $dV(X)$.

## 5. Amoeba's growth

The data set which we consider was experimentally obtained by Prescott in his excellent work[17]. The author also makes a note that similar growth curves were obtained for other organisms: "*The situation is strikingly similar to that described by Zeuthen ... for Tetrahymena.*" The growth curves were normalized to the *amoeba*'s initial weight. The growth scenarios were different. Most cells did not double their size, which is indicative that the nutritional environment was not sufficient for their normal growth or the dividing *amoeba* overgrew in the previous cycle. On the other hand, one *amoeba* significantly overgrew. Such diversity of growth scenarios is a good indicator that the nutritional conditions substantially varied during the growth of individual *amoebas*, as well as between different growth scenarios. Weighting procedures also contributed to variations in nutrient availability for individual *amoebas*.

However, in one case, an *amoeba* almost exactly doubled its weight (2.0504), which according to Ref. 5 means that it grew in a nutritionally *normal* and *stable* environment.

The argument in favor of stability is based on the fact that if not for the stability of the nutritional environment, the probability of exact doubling would be low.

**5.1.** *Finding the asymptotic value of the maximum possible volume*

Similarly to the approach used in Refs. 1-3, we model an *amoeba* by a disk whose radius increases while the height remains constant and equal to the initial disk radius. We assume that the cell's density is constant when an *amoeba* grows, so that we can substitute mass for volume. Ref. 17 supports this assumption as follows: "*This indicates that the density of the amoebae does not vary over the life cycle. Water uptake must be parallel to protein increase and to increase in dry mass*".

How adequate is the disk model of *amoeba*? *Amoeba* uses different nutrients, including smaller cells, for which it needs intensive endocytose. This is why *amoeba* needs a relatively large surface. *Amoeba* does not grow proportionally in all dimensions; this would quickly reduce its relative surface compared to relative volume, but *amoeba* needs surface to feed itself. So, the disk shape, when the disk height does not increase, is an adequate representation of this specific geometrical feature of an *amoeba*. The presence of pseudopods does not change this principal consideration. For instance, if we use a pinion-like shape whose height remains the same and the cogs increase proportionally in two other dimensions, the growth ratio curve will be nearly the same for the disk and for the pinion-like shape. The most definitive factor for the growth curve is in how many dimensions the disk changes. For instance, growth curves of disks with initial height of a half diameter and a diameter are almost indistinguishable. So, from the perspective of geometrical properties that are important for growth, the disk model is an adequate one.

Let us find the relative surface, the relative volume and the growth ratio according to (5) – (7). We need to know the maximum possible volume, which can be found using different methods. The maximum possible volume, in mathematical terms, is a horizontal asymptote $V_0 = const$, to which the growth curve defined by the growth equations (1) or (23) approaches when time $t \to \infty$. We find $V_0$ using the following method. For the last 3-4 points on the growth curves from Ref. 17, we calculate tangent of slope for the lines connecting the about equidistant experimental points. If points are not equidistant, then we can approximate location of such point using some smooth fit curve. (We can also use the points separated by one point between them, in order to reduce the impact of measurement errors.) It turns out that the series of values of these tangents (let us say for angles $A_1$, $A_2$, $A_3$, …) is approximately a geometric series, so that we could iterate the values of tangent further using the averaged common ratio $q = tgA_2 / tgA_1$ and compute the total vertical increment as a sum of infinite geometric progression $H_S = T \times tgA_3 /(1-q)$, where $T$ is the time period between two neighboring points. As an example, for three iterations, this approach produced the increments presented in Table 1.

Table 1. Finding the maximum possible weight using tangent iterations.

| tangent | 0.0388 | 0.021 | 0.0115 | 0.00626 | 0.0034 | 0.001 | 0.003 |
|---|---|---|---|---|---|---|---|
| increment | | | | | 0.02894 | 0.00984 | 0.00289 |

Summing up the increments and adding the value of $V_0 = 2.0504$ corresponding to the last measured volume, we find $V_0 = 2.0914$. (The beginning volume is equal to one.)

In [21], the parameter "spare growth capacity", *SGC*, was defined, which is interpreted as unrealized growth potential.

$$SGC = 1 - V_d / V_f \qquad (24)$$

where $V_d$ is the volume the division takes place at, and $V_f$ is the maximum possible volume.

It is interesting to note that the obtained number corresponds to the value of the spare growth capacity of about two percent. Close values, in the range of 1.0-2.8% have been obtained for other experimental growth curves (1%, 1.1%, 1.2%, 1.5%, 1.5%, 2.8%). The narrow range of values of the spare growth capacity is a good indicator that the actual cell division takes place at nearby values of this parameter. Also, we used the following independent approach to verify these numbers. Using the growth equation, we found the best fit of experimental data to computed growth curve when we assumed that we do not know the spare growth capacity. Then, using the obtained data fit curves, we computed the spare growth capacity. The obtained values matched the range of above numbers very well; all were within a range of (1.27 - 2.36%).

Some readers of the previous publications about the growth mechanism argued that once we begin to choose parameters of the growth equation, the growth curve could be adjusted to experimental data by manipulating these parameters. As we could see above, the methods of finding asymptotic value $V_0$ are quite objective and allow for only very minor variations, which would have a very insignificant impact on the shape of the growth curve. However, things are actually far more interesting and there are *fundamental* reasons for the introduction of the spare growth capacity parameter. Recall the citation from Ref. 4 about the remarkable conservation of the biochemical mechanisms across different species. The range of *SGC* of 1-2.8% that we found for *amoeba* reflects the arrangement of an *actual* growth and division mechanism. In Ref. 17, the author describes behavior of two *amoeba* that did not divide after the usual period of time and continued to exist in non-divided state for another 20 hours. These *amoebas* actually stopped growing at the usual division time (about 21-23 hours) and increased their mass very little, of about 2% (the mass of one *amoeba* then started to slowly decrease, according to experimental observations). This experimental data confirms the following. First, the spare growth capacity *really* exists, since *amoebas* continued to grow; although very little, but *exactly within* the predicted range. Second, our evaluation of the range of *SGC* is correct.

We also computed growth curves for *S. cerevisiae* using the value of *SGC* of 2%. Adjusted for specific growth features of *S. cerevisiae*, which replicates by budding, the correspondence of computed growth curve to experimental data from Ref. 18 was very good.

Given a great variety of growth conditions, it is common sense that the division takes place not at a certain value of *SGC*, but in the *range* of its values. However, it is also certain that this range is very narrow, roughly 1%. This fact again brings to light the saying from Ref. 4 about universality of major biochemical mechanisms, in particular the ones responsible for the cell division. So, we may formulate a credible hypothesis that *the actual grown size for organisms that exercise almost the whole growth cycle, like amoeba and S. cerevisiae, is less then the maximum possible size by the value of the spare growth capacity in the range of* $(1.9 \pm 0.8)\%$, possibly less. We may consider this value as a statistical characterization of a *real natural phenomenon*.

### 5.2. *Finding growth ratio and computing growth curves*

Besides the relative volume and surface and the growth ratio, we need to define the volume's differential for a disk, which is $dV(r) = \pi H((r+dr)^2 - r^2)$, and the disk surface $S = 2\pi \times r(r + H)$. As we found before, the nutrient influx for *amoeba* modeled by disk is defined by (21). Substituting these parameters into the growth equation (1), we compute the growth curve for *amoeba* whose size increases at 2.0504 and compare it to experimental data for a similar growth scenario. Note that we could use the form of growth equation (23), but then we should substitute the *minimum* specific influx, since the influx (21) takes into account nutrients required to support transportation and signaling networks. Fig. 2 shows computed growth curve versus experimental data.

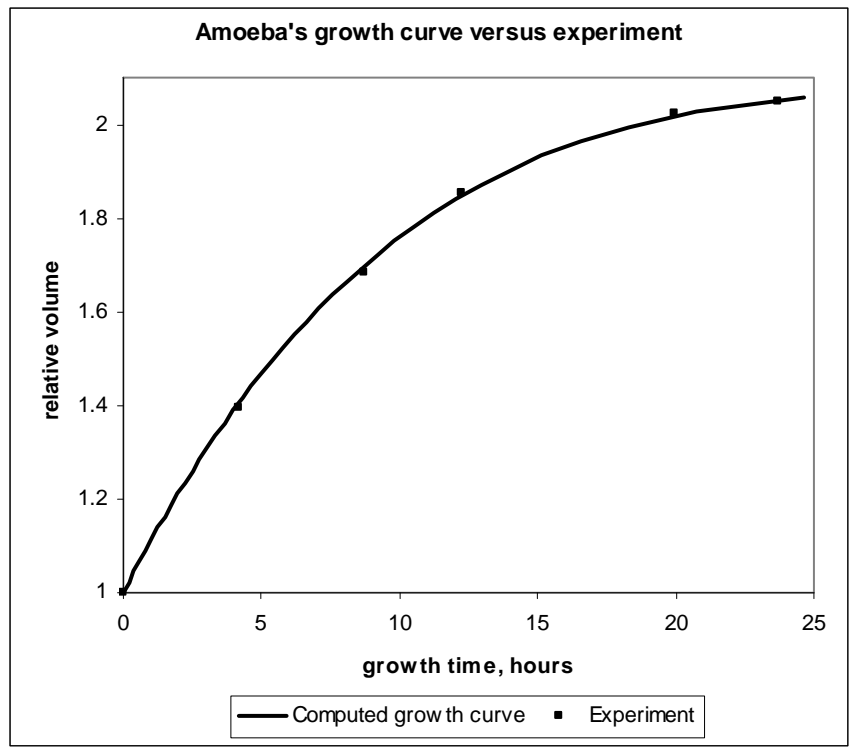

Fig. 2. Computed growth curve for *amoeba* modeled by a disk with height equal to initial radius, versus experimental data from Ref. 17.

We can see that the computed growth curve very accurately goes through experimental points. The only parameter that we adjust is the time scaling coefficient, which does not change the form of the growth curve. The value of the spare growth capacity is 2%. As we showed, this is not a data fit parameter, but a fairly universal characterization of growth of organisms that use almost the whole growth cycle. The rest of parameters of the growth equation has nothing to do with experimental data, but depends only on geometrical characteristics of *amoeba's* model and *amoeba's* biochemistry through nutrient influx. Overall, we have an excellent correspondence between the experimental data and the growth curve computed on the basis of *very general* considerations, which is a solid proof of validity of the general growth mechanism, growth equation, and our considerations regarding the nutrient influx required to support *amoeba's* maintenance, biomass synthesis and functioning of transportation and signaling networks.

Fig. 3 shows the growth curve for another set of experimental data, which corresponds to "underfed" *amoeba* (its weight at division is less than double). This curve presents one of the worst correspondences to experimental data. We can see that the computed growth curve fits experimental data fairly well. We used the value of spare growth capacity of 2.0%. Although the maximum possible size for the underfed *amoeba* might be higher, in which case the growth curve fits experimental data better, we did not speculate on that. The growth equation is applicable to all growth scenarios and certainly accommodates nutritional shifts during growth, whish is the case presented in Fig. 3, as we discussed before. However, we need more detailed information about such nutritional shifts and other variable factors in order to more accurately model complex growth scenarios. Later we will see how accounting for additional nutrient influx required for DNA synthesis improves accuracy of computed growth curves for wild type *fission yeast* and its mutant. From the mathematical perspective, the increase of nutrients influx in the pre-division period by few tens of percent, which is a realistic assumption, would result in the growth curve that accurately fits experimental data in Fig. 3.

Overall, the presented results provide valuable insights into many aspects of growth of living organisms. In particular, they are useful for the purposes of metabolic engineering, such as metabolic flux analysis and metabolic flux control, allowing to accurately specify nutritional and other conditions required for productive activity of industrial cultures in biochemical reactors.

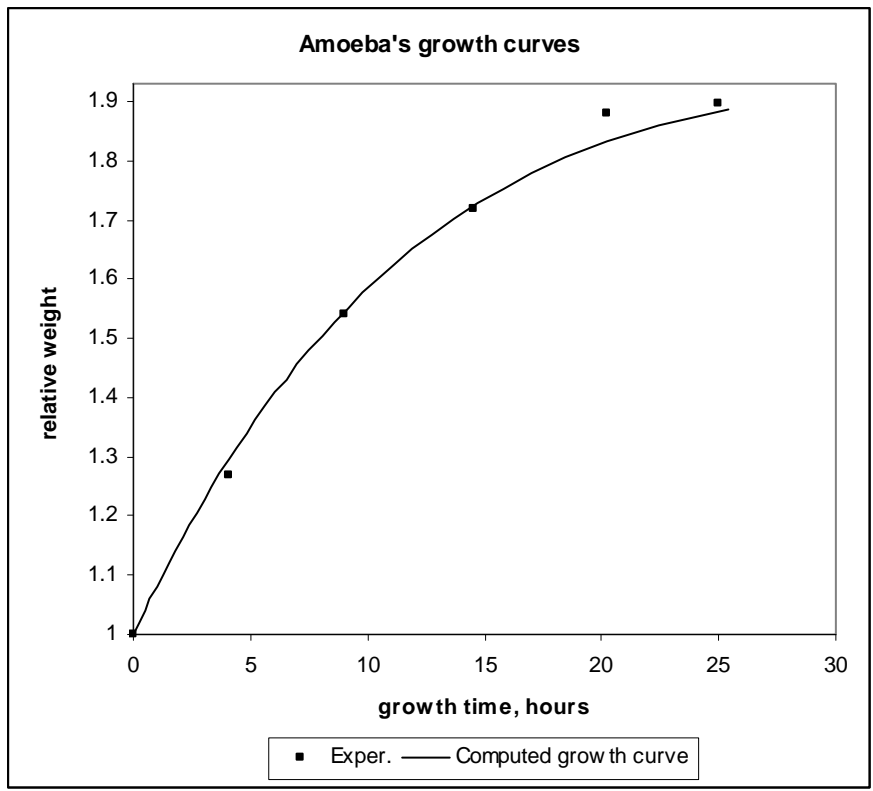

Fig. 3. One of the worst fits of the computed growth curve to experimental data from the series of experiments presented in Ref. 17.

**5.3.** *Amoeba's division mechanism*

In the previous sections, we studied the growth phenomenon. The authors of Ref. 6 refer to Refs. 19, 20 saying that they were able to achieve "*preventing Amoeba in G2 phase from entering mitosis indefinitely by periodically resecting a portion of its cytoplasm, thereby preventing the attainment of a presumed critical size*". So, conclude the authors, apparently, size matters for division. According to the physical growth mechanism, what matters first of all is not exactly the size, but the value of the growth ratio which indirectly depends on the size and the maximum possible size.

The growth ratio defines what fraction of the total influx is used for biomass production. It is known that the composition of biochemical reactions in a cell is very closely tied to the amount of produced biomass. Authors of Ref. 15 acknowledge: "*Sensitivity analysis indicated the importance of biomass composition on the model solutions*". In Ref. 15 the authors found for the metabolic flux analysis model that "*the model shows a high degree of sensitivity to the biomass information, and therefore the dependence of biomass composition on growth rates is an important aspect of a flux based model*". In the same work, the authors proved that small variations of biomass production in their model led to *significant* discrepancies with experimentally observed data. A similar high sensitivity of biochemical reactions to the quantity of produced biomass has been reported in many other works on metabolic flux analysis[16]. Earlier, we assumed that the growth ratio defines the amount of produced biomass. This approach proved to be correct with regard to modeling growth of *amoeba*. The logical inference from this would be that the amount of produced biomass, which in turn is defined by the growth ratio, defines composition of biochemical reactions.

Let us take a closer look at the specifics of *amoeba's* growth cycle before and during division. In Ref. 17, the author says the following, based on results of experiments: "*In each case growth of the amoebae virtually ceases about four hours before division takes place*". He also says: "*The transition to what I shall call the predivision period is not sharply defined*". The author also made an interesting note regarding the predivision period, saying that "(it) *is of special interest because it constitutes an unknown extending between the completion of growth and an onset of cell division. Its presence would seem to eliminate the consideration that cell size plays an immediate role in the initiation of cell division*". In other words, the author thinks that it is not directly the cell size, but rather something else that defines the transition to the division phase. The growth ratio is that factor that triggers the division.

Let us consider the behavior of experimental and computed growth curves in Fig. 2 and Fig. 3 in the last four hours. Table 2 shows the change of growth ratio depending on the relative volume and time. For the data in the last two lines, which correspond to 4.5 hours, we can see that the change of volume by 2% (which corresponds to experimental dependencies) causes a reduction of the growth ratio by 50.5%, which is a significant amount. So, although the volume (mass) of *amoeba* in the last phase of growth changes little, roughly by 2%, its growth ratio changes by *fifty* percent. Note that in the previous 2.7 hours the value of the growth ratio decreases by the substantial amount of 34%. This explains well the biochemical transformations in the predivision period described in Ref. 17.

Table 2. Absolute and relative change of growth ratio and volume depending on time.

| Time, hr | Volume | Volume change, % | Gr. ratio | Gr. ratio change, % |
|---|---|---|---|---|
| 12.743 | 1.857 | | 0.02186 | |
| 14.192 | 1.8968 | 2.14 | 0.017304 | 20.84 |
| 16.074 | 1.9369 | 2.11 | 0.012843 | 25.78 |
| 18.758 | 1.9775 | 2.1 | 0.0084734 | 34.02 |
| 23.426 | 2.0186 | 2.08 | 0.0041934 | 50.51 |

As we discussed at the beginning, according to the growth equation, the growth ratio defines what fraction of the total influx is used by the cell's biochemical machinery for biomass synthesis. Although the *amoeba's* mass does not increase much at the end of the growth cycle, the *rate of change of the growth ratio* is significant until the end of the growth cycle. This is an extremely important aspect, because the growth ratio determines how much nutrients from the incoming influx are used for *biomass production* $m_B$, as it follows from the growth equation (1). In order to see this in explicit form, we can rewrite the right part of (1) as follows.

$$dm_B = K \times \left( \frac{R_S}{R_V} - 1 \right) dt \tag{25}$$

Is the substantial change in the growth ratio the factor that governs the growth cycle and replication of *amoeba*? As we know, in G2 phase, *amoeba* can be prevented from entering mitosis *indefinitely* by resecting a portion of its cytoplasm. So, the attainment of some critical cell characteristic, apparently related to size, is important for *amoeba's* division. However, is it the growth ratio or size, or some other factor associated with the size that triggers mitosis? Let us summarize our findings that will help to answer this pertinent question.

° Attainment of some critical cell size is important in order to enter and complete mitosis.
° The actual mass of an *amoeba* changes very little in the final phase of growth, so it is very unlikely that such a small change that depends on many random factors can trigger such important events in the cell cycle.
° The fraction of the total influx of nutrients that goes towards synthesis of biomass is defined by the value of the growth ratio, according to equations (1) and (23). This growth equation is presumed to be correct, based on proofs presented in earlier published works[1,2,3] and also in our study that showed that the growth equation correctly describes experimental observations of *amoeba's* growth, when we incorporate known biochemical mechanisms into the model.

° Biomass synthesis is one of the key factors that defines the composition of biochemical reactions, so that substantial changes in biomass synthesis have to produce substantial, including *qualitative*, changes of biochemical composition and accordingly biochemical reactions.

° The value of the growth ratio changes by roughly 50% in the last phase of growth. This consequently leads to a *substantial* decrease in biomass production, which, accordingly, can *only* be achieved by a *significant* change in the composition of biochemical reactions. However, this is exactly what is required in order to start mitosis and eventually cell division.

The inference from the above summary, which reconciles the presented facts and considerations, is this. The growth ratio reflects the distribution of resources between maintenance needs and biomass synthesis. A drastic change in biomass production, which is a consequence of a substantial change in the value of the growth ratio (which reflects the change in distribution of nutritional resources due to interaction of supply abilities through the *surface* and demands for nutrients of *volume*), leads to a change of composition of biochemical reactions in such a way that mitosis starts. Biomass synthesis is a very important characteristic of the cell's biochemistry. In fact, this is one of the most critical factors that define the composition of biochemical reactions, as we learned before, based on the cited works. Although these works present results for cells other than *amoebae*, as the authors of Ref. 21 say, "*because the stoichiometry of cellular metabolism is well defined and variations among different cells are limited to a few reactions*", we can safely assume that the mainstream biochemical reactions, and consequently the amount of synthesized biomass as a factor that defines the composition of biochemical reactions, is as important for *amoebae* as it is for other cells. So, if the value of the growth ratio drops by 50%, and accordingly the output of biomass decreases by 50%, then it means that the composition of biochemical reactions has to change significantly too.

In the case of *amoeba*, this change of biochemical composition is substantial enough to trigger the transition to a new qualitative phase – to mitosis and subsequent division. Note that the same mechanism, that is the change of growth ratio, triggers all other qualitative phases of the growth cycle, such as G1, S, G2. Certainly, it happens in combination with and through biochemical transformations. Cells do not necessarily divide. Some cells, once they reach their maximum size, stop growth and become quiescent cells, such as muscle cells and neurons. In such cells, the change of growth ratio forces the transition to a different composition of biochemical reactions that make such cells quiescent. This is the case when the cells use the whole growth cycle defined by the growth equation, like in the case of *amoeba*. If cells switch to quiescent state at inflection point of the growth curve, then other biochemical mechanisms, in addition to the change of growth ratio, force such a transition. The same situation is with differentiated, specialized cells. In them, the change of growth ratio, with the addition of specific biochemical "transitional" mechanisms, triggers the change of composition of biochemical reactions, which begin to "build" differentiated cells.

The growth ratio is not a magic number, but an objective expression of unity and interaction between spatial characteristics of any organism – its surface, through which nutrient influx is delivered, and volume, whose existence and functionality this influx supports. It is also interaction, and to some extent competition, between the needs of two inseparable opposites: using resources to support existing biomass, and, at the same time, synthesizing new biomass by using existing biomass as a production facility.

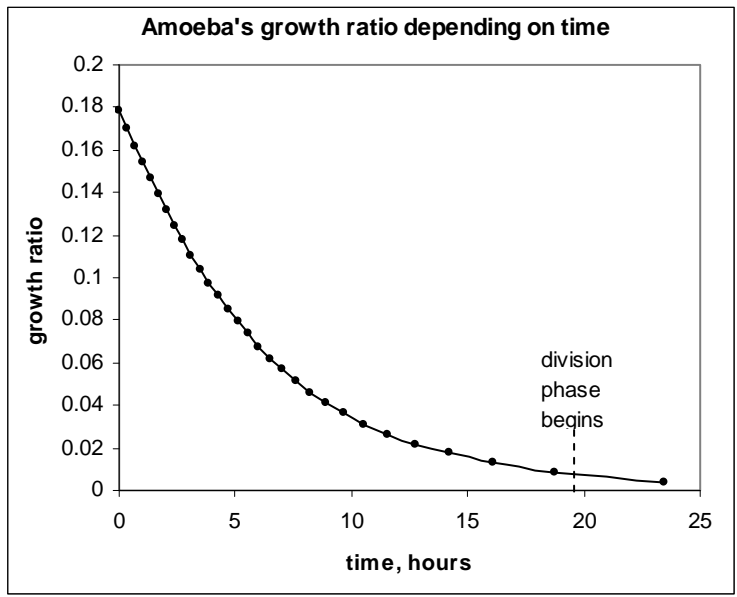

a

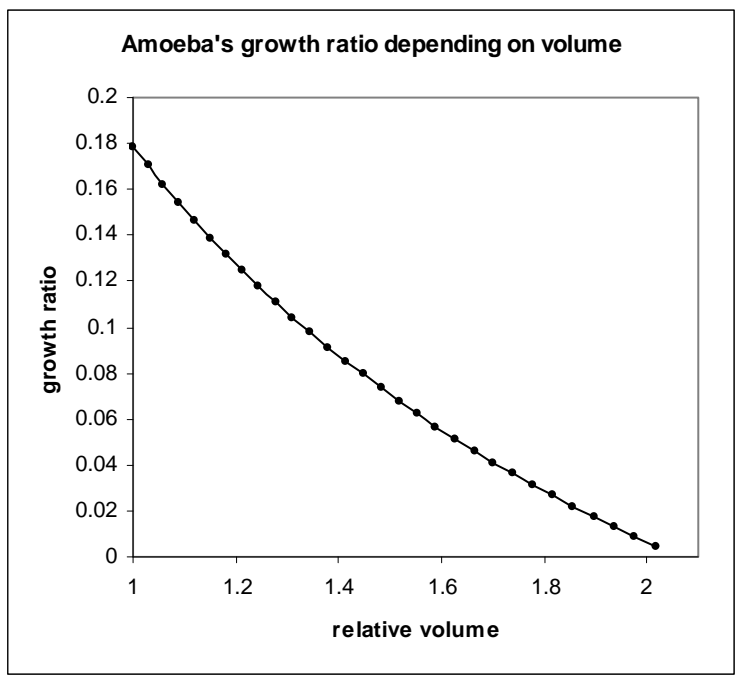

b

Fig. 4. Dependence of *amoeba's* growth ratio on time (a) and volume (b).

Fig. 4 shows dependencies of *amoeba's* growth ratios on time and volume. (We have to add to the said above that particular nutrient composition will be reflected in the composition of biochemical reactions as well, but it is still the value of the growth ratio that defines the composition of biochemical reactions, based on available nutrients.)

What about the change of growth ratio for experimental data? For the growth curves presented in Ref. 17, Prescott's *amoebas* during mitosis experienced a change of growth ratio in the range of 26-67%. So, the experimental data also confirm that the value of the growth ratio *significantly* changes in the last hours of *amoeba's* growth, although the weight and volume do not change much.

Change of growth ratio defines transitions between the known growth phases at the end of the growth period that lead to mitosis. Mitosis itself progresses at the lowest value of the growth ratio (0.0042 at the end versus 0.179 at the beginning of the growth cycle), so that the fraction of influx that goes toward biomass synthesis becomes negligible and mitosis is done mostly using the accumulated mitotic material, practically autonomously. Another confirmation of such a division and mitosis mechanism is mentioned in Ref. 17: "*Studies of the respiration of Tetrahymena show a leveling off of respiration shortly before division, indicating a cessation in the synthesis of respiratory machinery and probably, as in A. proteus, a general growth stoppage*".

As a side note, we should mention the following comment in the same work: "*During division in A. proteus phosphate turnover drops by 50%*". Recall that phosphorylation plays a crucial role in the activation and deactivation of mitotic material, so that the drop in phosphate turnover is a good indication that much of it binds to other substances[4]. This is what is required for activating CDKs that trigger disassembly of double DNA, as well as performing the remaining necessary actions in order to proceed through the subsequent mitosis phases and complete division.

The last note relates to the aforementioned statistical characterization of the spare growth capacity and its relation to change of growth ratio in the last growth phases. Since the growth ratio changes a lot, inevitable fluctuations of its value for organisms of the same type cause fluctuations of produced biomass and consequently fluctuations of composition of biochemical reactions, which explains variances in growth for organisms of the same type. How big these variances can be? We have no precise number, but based on the range of growth ratios for experimental growth curves, the amount of synthesized biomass at the end of division can differ by 2-2.5 times, which accordingly means quite a wide range of possible variations of compositions of biochemical reactions. This result agrees very well with known facts about flexible adaptation mechanisms developed in different cells[11] in order to secure replication. So, variation in biomass synthesis reflected in change of growth ratio is one of the mechanisms the adaptation mechanisms are based upon.

### 5.4. *Inherent unity of growth ratio, biochemical machinery and biomass synthesis*

It is important to understand that the synthesis of biomass is not a separate activity from the maintenance of existing biomass. There are no two separate production facilities in a cell, of which one strictly supports synthesis of biomass and another its maintenance needs. All biochemical reactions are related to each other, they are connected by numerous feedback loops, so that the change in composition of one branch of chemical reactions propagates through the whole cell's biochemical machinery. The synthesis of biomass, by and large, depends on *all* biochemical components and *all* biochemical reactions taking place in a cell. In order to change the amount of produced biomass, the *whole* biochemical machinery of a cell has to change. If the amount of produced biomass is reduced by two times (which happens in the last phase of the growth cycle of *amoeba*, when the growth ratio is reduced by 50%), then it means that the *whole* composition of biochemical reactions has to be *significantly* rearranged in order to accommodate this two fold decrease in biomass production.

An important mechanism developed by Nature is how much nutrients are allocated for biomass synthesis, and, consequently, how much biomass will be produced, which is tied to the value of the growth ratio. This is probably one of the most difficult postulates to accept. A natural reaction could be doubt, whether this relationship between a geometrical parameter and the biochemical machinery takes place. In order to overtake this threshold, those willing to understand the general growth law have to realize that the growth ratio is not an abstract geometrical parameter. The growth ratio is a mathematical expression of distribution of nutritional resources between maintenance needs and biomass synthesis at a given moment, in order to provide for the *optimal* growth of cells, without jeopardizing maintenance needs and without impairing biomass production, so that the cell could optimally proceed through the growth cycle. We found that the distribution of resources depends on the geometrical properties of organisms; a fact to which we should not object, because, indeed, resources are supplied and distributed in a *space*, and hence the geometrical properties of this space have to influence the delivery and distribution of resources. (For instance, life organization and flow of resources between numerous dispersed small towns along a single road, let us say in a mountain valley, very much differs from life of the same number of people in a compact high-rise city just because of different geometrical organization of their living space.) Of course, criteria of optimal distribution can be different depending on the particular organism. Apparently, for most organisms, the criterion of fastest growth and reproduction is most common, although other criteria are also conceivable.

**5.5.** *Some inferences from the growth equation*

Besides the accurate modeling of growth, the general growth and replication mechanism explains numerous issues related to growth phenomenon. In Ref. 3, the new growth suppression mechanism based on change of organism's geometry was discovered and confirmed by experimental data for Drosophila and pigs. Here we consider a few more examples.

5.5.1. *Why the growth rate is the highest at the beginning?*

This question was formulated in Ref. 6 among others pertinent questions related to growth. It is known that the growth rate at the beginning is the highest. In Ref. 17, the author acknowledges: "*There is no initial lag in the course of growth of an amoebae following division; on the contrary, its initial rate of increase of weight, volume, or protein content is its highest*". In another place in the same work the author notes: "*With respect to the growth phase for the whole cell the rate of synthetic activity is the highest directly following division but falls off slowly until the predivision period is reached.*".

 Now, we can confirm these experimental findings by computations based on the growth equations (1) and (23). Let us take a look at Fig. 4. We found that biomass production is defined by the fraction of influx that is used for biomass synthesis. This fraction, in turn, is defined by the growth ratio. The graphs of growth ratios presented in Fig. 4 confirm that the rate of biomass synthesis is *highest* at the beginning of the cell cycle because the value of the growth ratio is the highest at the beginning. This is what the authors of the aforementioned work observed in their experiments. So, our results are in full agreement with these experimental observations and explain this feature of growth phenomena very well.

*5.5.2. Why can the same cells grow small and large?*

Another interesting question asked in Ref. 6 is why cells (or organs or whole multicellular organisms for that sake) of the same type can grow larger or smaller. The general growth mechanism explains this secret of Nature as well. As we discussed earlier, large cells have excess of ribosomes compared to what they need for protein and RNA synthesis, so that the actual number of ribosomes that is actively involved in synthesis can be as low as about 10%. Ribosomes are the last cell components that degrade when cells starve. These observations are very well confirmed by experiments with nutritional shifts[11]. Once the culture was moved into richer in nutrients environment, the rate of growth increased almost immediately, at least no experimental observations could discover any delay. This means that normally cells, unless they are starved almost to death, maintain biochemical mechanisms that allow immediately processing more food once the cells are moved into richer in nutrients environment. This reserved processing capacity is large, as experiments described in Ref. 11 show. Such an ability of cells to increase their processing capacity by several times when nutrients are abundant (which requires an appropriate increase of influx) was developed evolutionarily, which is a very natural adaptation mechanism.

In *relative* terms, the fraction of nutrients directed toward biomass production is defined by the value of the growth ratio. When nutrients are abundant, the fraction of influx for the biomass synthesis in *absolute* terms (which is equal to influx multiplied by the growth ratio according to equation (1)), can be substantially, several times, larger in cells growing in richer in nutrients environment. So, according to the growth equation, in a rich medium, the cell's biomass increases faster.

The composition of biochemical reactions at the beginning of growth is largely the same for cells that would later become small or large. However, when the size increases, a bigger cell's mass requires additional resources compared to a cell with a small mass. The difference is due to different energy and nutrient distribution in small and large cells and some biochemical mechanisms specific for the support of larger cells. This is one of the reasons why the ribosome content is higher in larger cells[11]. It follows from the above that maximum possible size may change during growth. If we assume the same nutrient availability throughout the whole growth cycle, then the evaluation of maximum possible size can be done in early growth stages, such as during DNA synthesis, based on features of biochemical reactions specific to a particular cell, characteristics of nutrients, and their concentration. However, if the nutritional medium is shifted, then the growth scenario changes too, depending on the growth phase[11]. The shift will unlikely affect the growth of cells in the mitosis phase, but it can change the cells' biochemical composition and composition of biochemical reactions in the previous growth phases, thus changing the value of the maximum possible volume for this particular cell. In Ref. 11, the authors say the following with regard to *E. coli* culture subjected to a nutritional shift: "… *the average length and the thickness of the cells begin to increase immediately after the shift*". They also found that "*it seems the cells are capable of switching almost instantaneously to a* *greatly increased rate of RNA synthesis, and to maintain this new rate for a period long enough to double their initial RNA content*".

So, there are some specifics of biochemical composition of larger and smaller cells of the same type. These specific features form during growth. The differences result in increase of the maximum possible size for cells growing in nutritionally rich environments. This difference may not exist or be small at the very beginning, but during growth, the larger cells, due to availability of nutrients, create internal structure that supports bigger maximum possible size (and consequently grown size).

What would be an optimum growth scenario for bigger cells compared to small ones? Still, we have the same problem of optimizing the distribution of available resources between maintenance and biomass synthesis. However, according to the general growth law, this optimum at any given moment depends on geometry of organisms and composition of biochemical reactions via the maximum possible size. So, we should expect that growth of small and large cells of the same type with the same geometry in non-stressful conditions should be *invariant* to the initial size. This is like all growth processes are scaled up for larger cells. In particular, it means that the growth time should be about the same for small and large cells, unless the growth conditions are such that biochemistry of organisms changes. (For instance, this happens when organisms starve.) A remarkable experimental confirmation of existence of such a scaling of growth mechanism can be found in Ref. 22 (see Fig. 1B from this work), in which authors present measurements of growth cycle time for three organisms: wild type *fission yeast*, *fission yeast* mutant that grows small, and diploid wild type *fission yeast* with a large size. The average growth time is *the same* for all three organisms, despite their significant difference in size (three fold). Note that this scaling is fairly precise in spatial terms as well, that is both initial and final lengths are scaled by three fold too (these organisms grow only lengthwise).

This scaling of growth also confirms validity of our earlier assumption that nutrient requirements for supporting transport and signaling networks are proportional to distance. If it was not the case, then bigger cells would most likely require either disproportionably more or disproportionably less nutrients in order to support their transport and signaling networks, and accordingly they would grow slower or faster. However, as we found out, this is not the case.

Now, what about the extreme growth scenarios? Strains of some organisms exhibit relatively small variability in the length of G2 phase, and generally the lengths of S/G2/M phases are significantly more stable[6] than the length of G1 phase. Why is it so? The reason is that before proceeding to S phase cells have to create a "biochemical plant" of certain predefined capacity in order to synthesize DNA. This process cannot be scaled down, since the DNA cannot be scaled. So, too small or starving cells first have to build such an internal infrastructure that would allow to begin synthesis of DNA, which requires an extended period of time, when nutrients are low or the initial size of the cell is small. This is why the length of G1 period is more volatile than the lengths of other growth phases. Accordingly, this period is characterized somewhat different biochemistry compared to cells that were born at normal size and begin synthesis of DNA at the very early stages. These *qualitative* differences in biochemistry of small cells at initial growth phase, compared to normal or large cells, create divergences from the straightforward scaling of growth for small and large cells. Otherwise, the growth of small and large organisms of the same type is largely the same and can be described using a scaling factor in a wide range of sizes. Certainly, other factors can interfere and impose some limitations, like physical ones. For instance, bones of exactly the same shape and structure but of different sizes do not have scaled physical properties, such as the maximum possible load when bones break.

So, the size of a grown cell is defined by its geometrical characteristics, specific composition of biochemical reactions and available nutrients. Geometry and biochemistry of an organism are interdependent and influence each other. In turn, the biochemistry depends on nutrients. In fact, nutrient availability influences geometrical characteristics too. For instance, in a richer medium, cells can resume growth by elongating; this effect was explained in Refs. 2, 3.

Two illustrations presented in this section give an idea of how to approach different pertinent issues of cell growth. In particular, the last paragraph allows explaining the differences in the length of different growth phases, depending on the nutritional medium and initial mass of the newly born cells. It also helps explaining why small cells, which grew for several generations in a nutritionally poor medium, restore their normal size when they are moved to nutrient-rich environments. The fact that sometimes such a nutritional shift requires several generations of cells to have an effect on the cells' maximum size, confirms very well our inference that the compositions of biochemical reactions of small and large cells of the same microorganism are somewhat different. This is why the biochemical transition to the composition corresponding to larger cell requires some time. How long, depends on the particular organism and how drastic the nutritional shift is.

**6. The fastest growth scenario**

**6.1.** *Inflection point of the growth curve as a trigger of division*

In this section, we model the growth of *S. pombe*. This organism presents a different type of growth, whose growth scenario drastically differs from the steady, asymptotic-like growth of *amoeba* and other similar organisms. There are other examples of fast growth scenarios, such as the growth of *E. coli*. In this case, cells switch to division phase when they reach an *inflection* point on their growth curve. From the evolutionary perspective, this is a very logical arrangement, because it provides the *fastest* possible growth among all other scenarios. If the switch to the division phase would be slightly delayed, to occur *after* the inflection point, then the growth time would be larger. This is less optimal from an evolutionary perspective. Similarly, if growth stops *before* the inflection point, then the time needed to get to the same mass would also be larger. Maximization of growth rate (in other words, biomass production) is the optimization criteria which is successfully used in metabolic flux analysis[16], which confirms the importance of this parameter for the growth cycle.

The value of the growth ratio still defines the fraction of the total influx directed toward biomass production and, through the quantity of produced biomass, the composition of chemical reactions. However, this time, unlike in *amoeba*, switching to the division phase is tied to the point where the decrease of growth rate begins. Two factors, that is, the value of the growth ratio and the decrease of the growth rate, are interdependent, but the decrease of the growth rate leads the game. Earlier switching to division phase is supported by evolutionarily newer more sophisticated biochemical mechanisms. In particular, in fission yeast, there is a mechanism that creates spatial gradient of cyclin kinase Pom1 in such a way that its concentration increases at the ends of a cylinder like fission yeast and decreases at cortical nodes, which triggers a chain of phosphorylation based reactions starting division[23,24].

Mathematically, the maximum of the growth rate corresponds to the inflection point of the growth curve. Fig. 5 shows a graph of the growth rate, which is the first derivative of the growth function, depending on time for a particular growth scenario of wild type *fission yeast*.

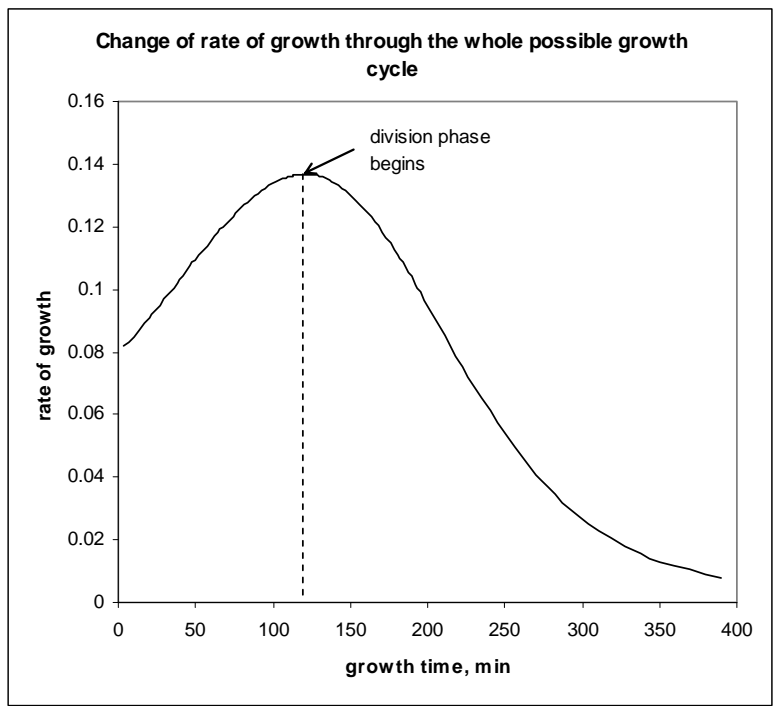

Fig. 5. Change of growth rate for the wild type *fission yeast*. The growth curve for the *whole possible* growth cycle is shown.

Fig. 5 shows the change of rate of growth for the *whole* possible growth scenario, as if the cell continued to grow after the inflection point, until reaching a maximum possible size. In Refs. 4, 6, 16, 17, we can find examples when certain cells were forced to pass the start of division phase at the inflection point of the growth curve, and continued to grow larger than their usual size, so that the possible growth curve is *real*.

What would be the spare growth capacity of such cells? It depends on the initial size, nutrition and composition of biochemical reactions. In case of example in Fig. 5, the maximum possible length is about 2.76 of the initial cell length. The value of the spare growth capacity $SGC \approx 34.5\%$.

Fig. 6 presents an example of a growth curve for the *whole possible* range of growth of *S. pombe*. We can see that it is very similar to what we discovered for *amoeba*, although in that case it was the actual growth curve. However, if we suppress the division of *S. pombe* at inflection point, which is possible, then the curve in Fig. 6 will become an *actual* growth curve.

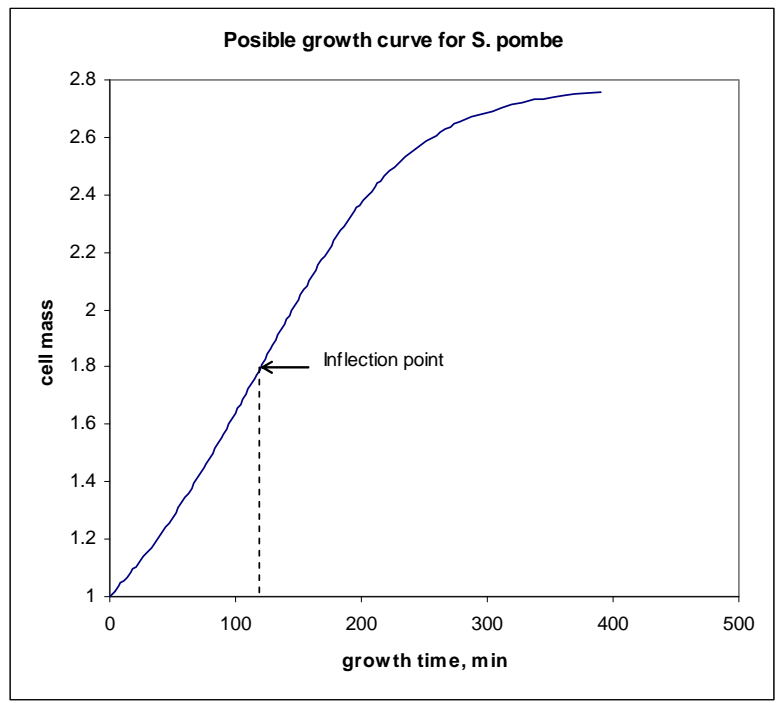

Fig. 6. Possible growth curve for *fission yeast*.

The dependence of the growth ratio on time for the whole possible growth cycle is shown in Fig. 7. We can see that this dependence is similar to the one for *amoeba*.

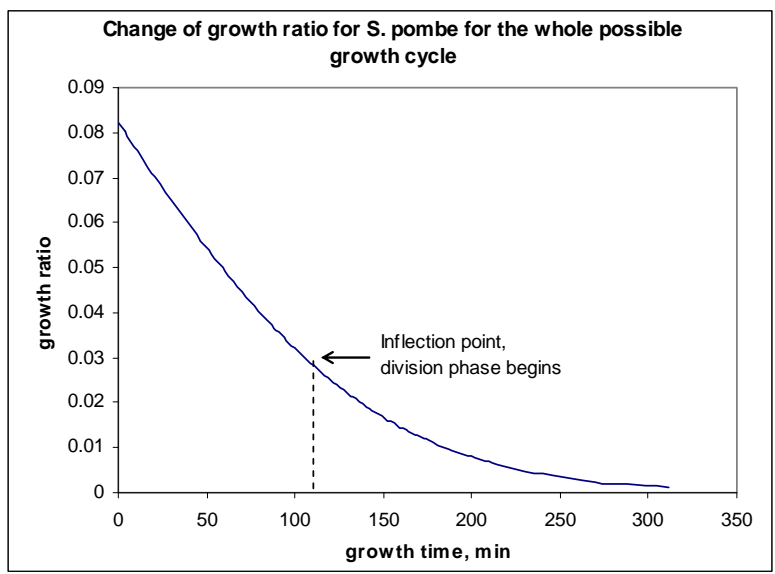

Fig. 7. Growth ratio of wild type *fission yeast* for the whole possible growth cycle.

### 6.2. *Modeling the growth of fission yeast*

Below, we model the growth using the growth equation (23) and compare results with experimental data for *Schizosaccharomyces pombe* presented in Refs. 22, 25. The cell is modeled by a cylinder with hemispheres at the ends. The model is able to incorporate the increase of both length and diameter, although, for *S. pombe*, we will consider the increase of length only, as its diameter remains the same during growth. In the case of *E. coli*, cells might increase both in diameter and length[15]. This observation is applicable to individual cells that grow in different nutritional media. If the *E. coli*'s diameter depends on the growth rate, then it also means that when the medium is shifted in terms of nutrient conditions, the cell has to grow thicker. In Ref. 11, the authors confirm this: "*With respect to the shape and dimensions of the cells, it should be noted that the big, fast growing cells are thicker as well and longer than the small, slow growing cells*".

We consider different chemical compositions associated with the growth rate in order to cover the *whole range* of possible variations. We do not have detailed data on chemical compositions for *S. pombe*, but we can substitute similar data for *E. coli* presented in Ref. 11. Based on available data, the protein and ribosome contents in *E. coli* and *S. pombe* are close. Moreover, because we cover the *whole range* of possible changes of ribosome and protein content, some minor variations, if exist, are not meaningful for our purposes. The important thing is to associate the *range* of possible growth rates with the appropriate chemical compositions. The summary of data that we use in calculations is presented in table 3.

Table 3. Chemical composition of *E. coli* cells, [mg/(g dry weight)], from Ref. 11.

| Scenario No. | $\mu$ [1/$hr$] | DNA | RNA | Protein+ tRNA+ | $C_R$ | $C_P$ |
|---|---|---|---|---|---|---|
| 1 | 0.2 | 40 | 35 | 915 | 0.035 | 0.924 |
| 2 | 0.6 | 37 | 90 | 870 | 0.09 | 0.873 |
| 3 | 1.2 | 35 | 135 | 825 | 0.136 | 0.829 |
| 4 | 2.4 | 30 | 250 | 730 | 0.246 | 0.723 |

We will substitute the values of $C_R$, $C_P$ into (20) in order to find the total and specific influx. We also need to compute the relative volume, the relative surface, and the growth ratio using (2) – (4). Then, we will substitute these values into the growth equation (23) and solve it relative to mass (which is proportional to volume when the density is constant), considering it as a function of time. Geometrical dimensions of the general model are shown in Fig. 8. (In computations for *S. pombe*, we will assume that the *cylinder's diameter does not change* during growth, that is $R = r$.)

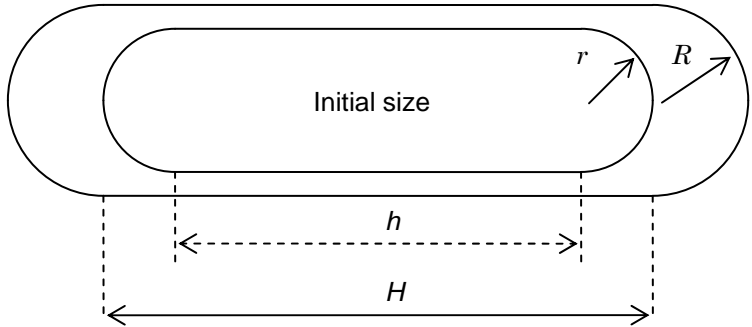

Fig. 8. Notations for the *E. coli*'s and *S. pombe*'s growth models.

For the notations in Fig. 8, we find.

$$R_S = \frac{r(2r+h)}{R(2R+H)} \tag{26}$$

$$R_V = \frac{r^2((4/3)r+h)}{R^2((4/3)R+H)} \tag{27}$$

$$G_R = \frac{R_S}{R_V} - 1 = \frac{R((4/3)R+H)(2r+h)}{r((4/3)r+h)(2R+H)} - 1 \tag{28}$$

Using (20), we can define the value of specific influx either using the cylinder's length,

$$k(R) = \left(C_{pL}(h/H)^2 + C_{rL}(h/H)^3\right)/S \tag{29}$$

or, using the whole length, as follows.

$$k(R) = \left(C_{pL}((h+2r)/(H+2R))^2 + C_{rL}((h+2r)/(H+2R))^3\right)/S \tag{30}$$

The cell's volume increases due to the elongation of the *cylindrical* part of the cell *only*, while hemispheres remain the same, they are just pushed sideways by the growing cylindrical part. So, in geometrical terms, all volume increase is done through the increase of biomass of the cylindrical part, so that influx for RNA and protein synthesis should be associated with the cylindrical part too. The same is true for the signaling and transport networks. This is why using (29) seems more adequate.

We also considered the case when influx is defined by (30), for the sake of completeness, but, as it was expected, using only cylinder's length in defining influx fits experimental data better. When we consider the cell's length, we certainly include hemispheres, because experimental data are based on dependence of the *whole* length on time.

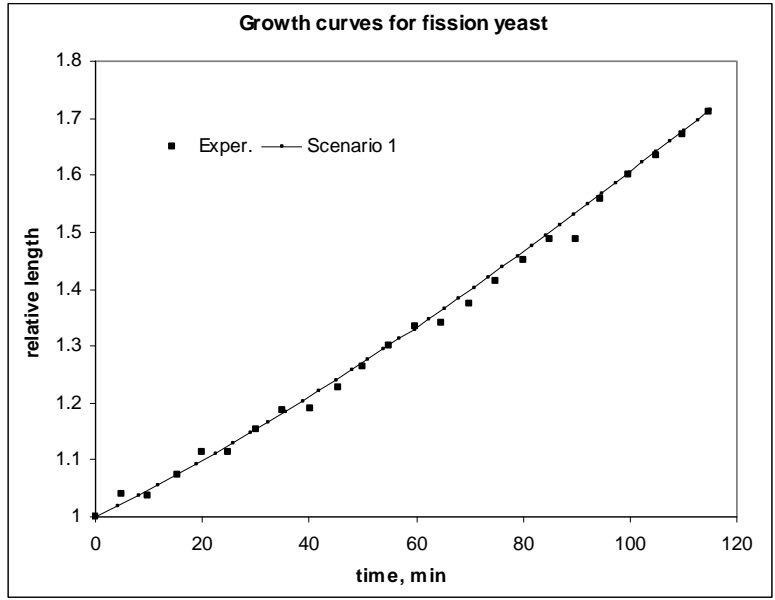

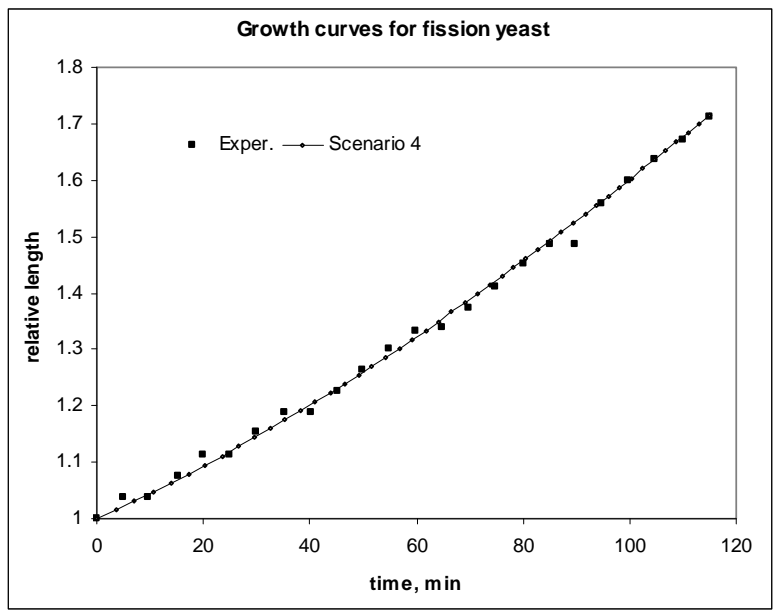

Fig. 9. Growth curves for different growth scenarios from table 3 based on data from Ref. 11 versus experimental data from Ref. 25.

Substituting (26) – (29) and data from table 3 into the growth equation, we obtain the following growth curves shown in Fig. 9. We can see that growth curves are very close and fit experimental observations well. The growth curve that corresponds to the highest growth rate has the highest curvature (convexity) (this is scenario 4; nutritionally, this is the richest medium), while the growth curve for the nutritionally poorest medium has the least curvature (scenario 1). This fact agrees very well with experimental observations that cells growing in poor media have *more linear* growth curves[4].

Note that all curves can be approximated by a broken line that consists of two linear segments, as the authors of Ref. 22 did. The difference in tangents of slopes for such lines, as the authors of the cited work found, was about 30%. About the same values, in the range of $\pm 2\%$, were obtained for the curves in Fig. 9 when they were approximated by a broken line with one break. The highest value was 32.2% (we mean difference in tangents of slopes) for the cell which grew in a nutrient-rich medium (scenario 4). This fact further confirms the validity of the growth equation.

### 6.3. *Change of nutrient influx during growth*

Influx dependencies on time and length for some of the considered growth scenarios are presented in Fig. 10. Compared to the case of *amoeba*, influx in the case of *S. pombe* increases faster than linearly for all three types of nutrient influxes (specific influx, total influx and influx per unit of volume). This reflects the increasing nutritional needs of a cell and accordingly the growing processing capacity of a unit of volume. Note that the increase of specific influx means that the same membrane surface is able to transfer more nutrients inside and accordingly more waste and materials outside. This increase of influx through surface is facilitated by the fact that the rates of increase of surface and volume for elongating geometrical forms are much closer than in case of a form that increases in all dimensions, so that the amount of volume per unit of surface does not increase as fast as, let us say, for a spherical form. So, the growth ratio for a cylinder-like cell imposes fewer restrictions on the increase of specific influx. These revealing findings confirm very well the fact that organisms such as *E. coli* and *S. pombe* are among the fastest growing organisms. During their evolution, they used every opportunity to optimize their development in a way that supports the fastest possible growth. This relates to tuning up their biochemical mechanisms to realize the fastest part of the growth curve and then switch to the division phase. Another adaptive feature that contributes to faster growth of these organisms is their cylinder like form. It was shown in Refs. 1, 3 that organisms that have a cylindrical form have the fastest growth time among elongated forms. (Besides, a cylindrical form also provides mobility, which is also apparently very important for such organisms.)

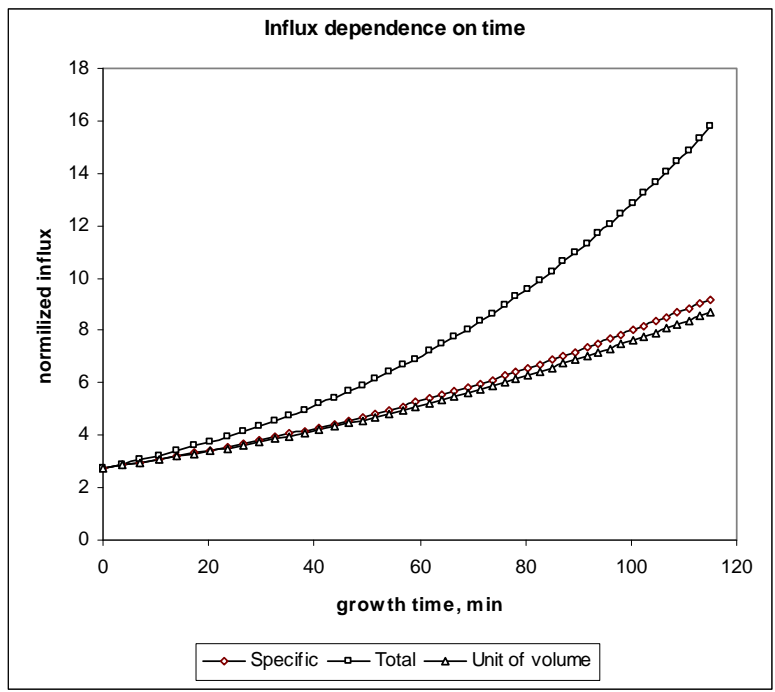

Fig. 10. Dependence of specific influx (amount of nutrients per unit of membrane surface in one unit of time), total influx, and influx per unit of volume on the growth time. All values are normalized to the same initial value.

### 6.4. *Modeling the growth of wild type fission yeast. Influx for DNA synthesis*

In this section, we use more accurate growth measurements from Ref. 22 for the wild type (WT) *fission yeast*. If one looks at Fig. 9, he may notice that at the beginning experimental points apparently go slightly higher than the growth curve predicts. However, the overall dispersion of experimental data from Ref. 25 did not allow us to reliably isolate this effect. More accurate measurements done in Ref. 22 allow locating this effect with more certainty. Like in the previous section, the inflection point of the growth curve corresponds to the point where experimental dependence sharply bends and the division phase begins. If we take into account only the influx required for protein and RNA synthesis, then the growth curves will look like the ones presented in Fig. 11. We show two scenarios from Table 3 for comparison, but the curves for all growth scenarios are very close. We can see that in the first fifty minutes the growth curve goes below the experimental points. This effect is due to DNA synthesis which our computed growth curves did not account for. Accordingly, we took into account only the influx that is required for protein and RNA synthesis.

In Ref. 8, the authors note that "*DNA synthesis is periodic with a peak of synthesis occurring between 0.1 and 0.4 of the cell cycle*". This specific feature is somewhat similar to what we observe in *amoeba*, whose DNA synthesis is also periodic, with apparently two phases: one is in the first five hours when about 75% of DNA is synthesized, and another in 9-13 hours, from the total growth period of 36-48 hours (at temperature of $23^0 C$). So, the DNA synthesis period in *amoeba* is slightly less than 0.2 of the growth cycle. If we take into account that the whole possible growth cycle for *S. pombe* is about two times larger than the actual one when the division phase starts at inflection point, then the DNA synthesis period for *S. pombe*, relative to the *whole possible* growth cycle, will also be about 0.2. So, when we consider the DNA synthesis cycle relative to the *whole* possible growth period, there is more similarity in the durations of the DNA synthesis cycle between these two organisms than it seems at first glance.

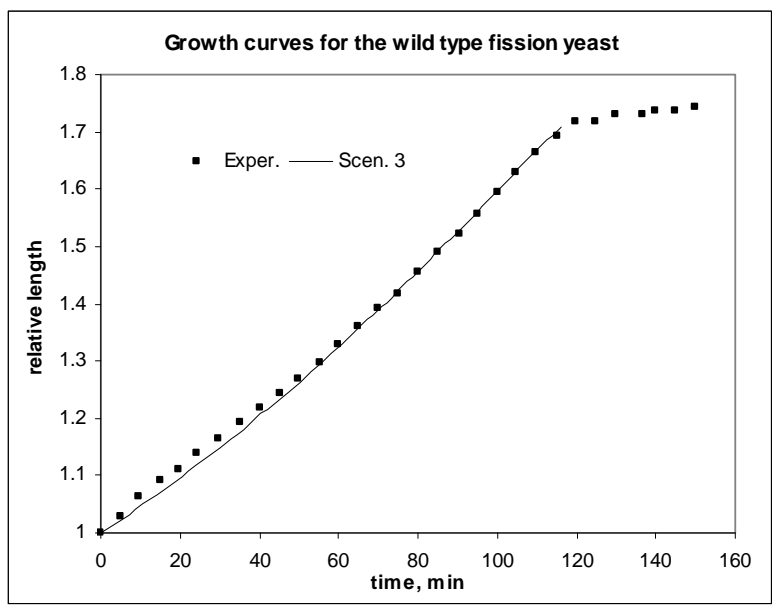

Fig. 11. Growth curves for wild type *fission yeast* versus experimental data from Ref. 22 for growth scenario 3.

However, these results are valid for cells that grow in nutritionally normal environments. When a cell starves, the DNA synthesis scenario can be quite different. It was noted in Ref. 6 and the other literature, that small cells have a substantially longer G1 growth phase than normally developing cells. This is because the cell has to grow to a certain size before DNA synthesis can start. The reason is that the biochemical machinery with appropriate synthetic capacity to support DNA synthesis has to be created first. The capacity of biochemical machinery required for DNA synthesis cannot be scaled much, as we discussed before, because the DNA itself cannot be scaled. We need a certain *absolute* amount of components and energy in order to duplicate the DNA, not more, not less. This is like a self-sufficient plant that has to manufacture a ship that has a length of one hundred yards and other fixed characteristics. The plant cannot spend less material than what was defined in specifications, and neither can the plant build a bigger or smaller ship. All sizes, materials, and production capacities have to match rigid specifications. In our case, until the cell builds the biochemical "DNA production plant" of a strictly predefined production capacity, it cannot begin DNA synthesis. This is why smaller cells have a substantially longer G1 period.

It is known that the energy and material consumption is highest at the center, at the nucleus, because this is where the cell doubles the DNA, which requires a lot of materials and energy for synthesis. Besides this, many more activities such as transcription are performed inside the nucleus. Another factor that contributes to higher energy consumption in DNA synthesis is that macromolecules require more energy for synthesis than simpler proteins. So, we may readily expect that the influx directed towards DNA synthesis is somewhat greater than a simple percentage of the DNA mass in the cell's dry weight. Still, we do not know exactly how much greater it can be, but putting an upper bound at no more than a few times greater would be a reasonable assumption.

The growth curve with adjustment for the influx required for DNA synthesis is shown in Fig. 12, while Fig. 13 shows influxes. For this scenario, the value of influx for the DNA synthesis is about 9% of the total influx on average, which seems like a reasonable value, compared to 3-4% of the dry weight corresponding to DNA mass. Further increasing the value of influx for DNA synthesis improves the fit to experimental data.

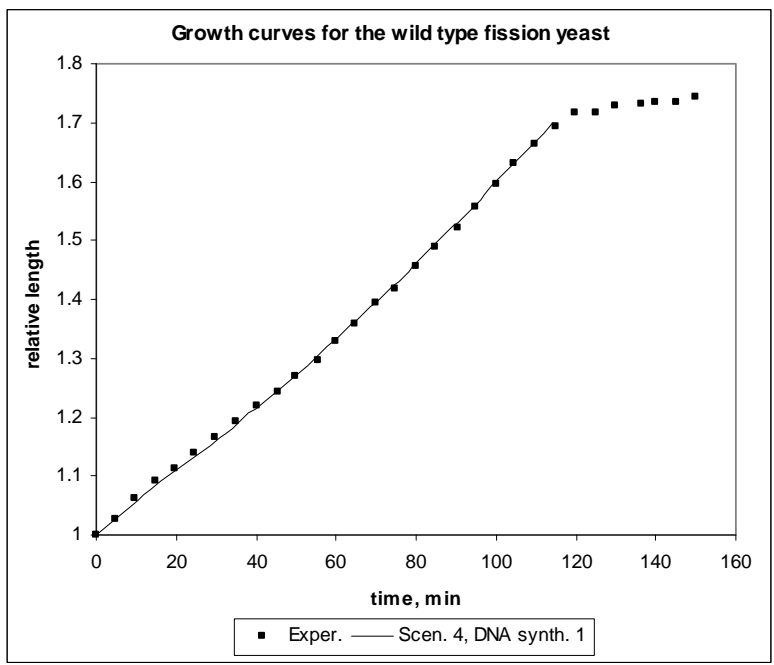

Fig. 12. Growth curve adjusted for DNA synthesis, when the influx increase is about 9%.

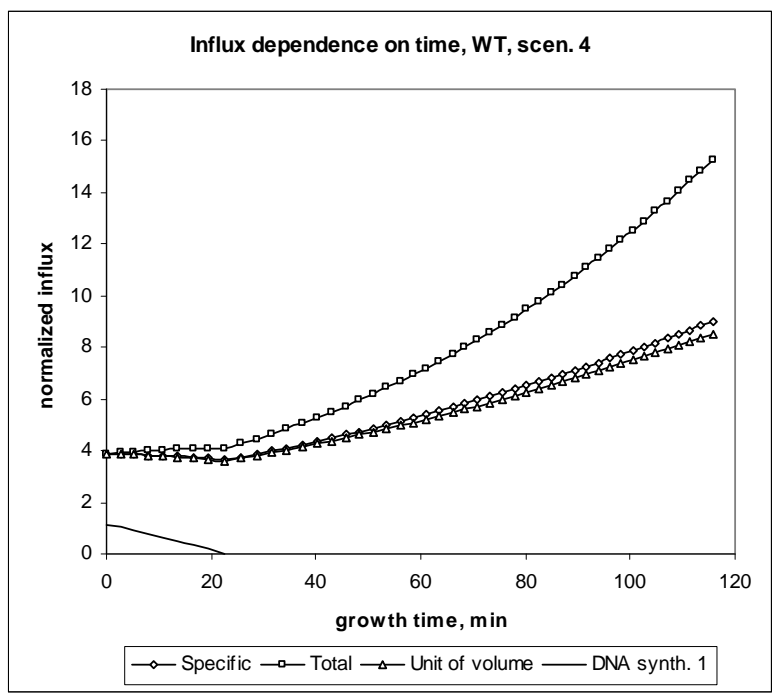

Fig. 13. Dependence of influx per unit of volume, specific and total influxes, and additional influx for DNA synthesis on time.

We would not speculate on what functional dependence of the additional influx for DNA synthesis is more accurate, which could be a subject of separate study. What is important at this stage is that the introduction of the reasonable additional influx significantly improves the model fit, and that the amount of required influx is commensurate with the high energy requirements of DNA synthesis. The approach has a practical value with regard to finding the *actual* influx required for DNA synthesis. Mathematically, this is a challenging but in some instances feasible task if the kernel of the integral growth equation derived from (1) or (23) possesses certain properties. Another option would be to find some properties of the additional influx based on nature of the growth phenomenon, or from experiments, and then parameterize the specific and additional influx dependencies, which might provide the required solution. In some cases, the growth equation has analytical solutions[3], which may also facilitate finding an explicit functional dependence of specific, additional and / or total influx on time or on volume.

Another aspect that is worth noting is that the change of slope for the obtained growth curves, when they are approximated by a broken line, corresponds very well to experimental value of 30% obtained in Ref. 22. Table 4 presents results of computations. Since experimental data corresponded to average nutritional conditions, we could expect that our most adequate computational scenarios would be 2 and 3. As we can see from the table both produce values close to the experimentally obtained 30%.

Table 4. Change of slope for the growth curves.

| Scenario | 1 | 2 | 3 | 4 |
|---|---|---|---|---|
| Change, % | 28.3 | 29.8 | 32 | 34.6 |

### 6.5. *Modeling the growth of* $wee1\Delta$ *mutant*

In the same work[22], the authors studied the growth of *S. pombe's* $wee1\Delta$ mutant. This mutant is about 1.66-1.69 times smaller in length than wild type *fission yeast*. Let us denote this scaling value as $s_{Wee1}$. The $wee1\Delta$ mutant divides earlier, when the cell is still small. We assume that except for the mutated Wee1 CDK functionality, the rest of biochemical reactions remain the same, so that this mutant has the same biochemical machinery that can potentially allow it to grow to a normal size. This is a reasonable assumption since no other differences in biochemical properties were reported. In order to reach the normal cell size, this mutant, whose maximum length we denote as $L_{mut}$, would have to grow to the size $L_{normal}$. So, $s_{Wee1} = L_{normal} / L_{mut}$.

The inflection point on the *S. pombe's* growth curve corresponds to the size at division of the wild type grown cell. We can compute the location of this inflection point in units of initial volume of the mutant using the value $s_{Wee1}$. The value of such defined relative volume, corresponding to inflection point for the mutant had it been allowed to grow to the size of wild type *S. pombe,* can be found as $V_{\inf Mut} = V_{MutGrown} \times s_{Wee1}$. In our case, the relative volume of the grown mutant according to experimental data is about 1.69, so that

$$V_{\inf Mut} = V_{MutGrown \approx} \times s_{Wee1} = 1.69 \times 1.68 \approx 2.85$$

The graph of the growth ratio for the mutant is shown in Fig. 14.

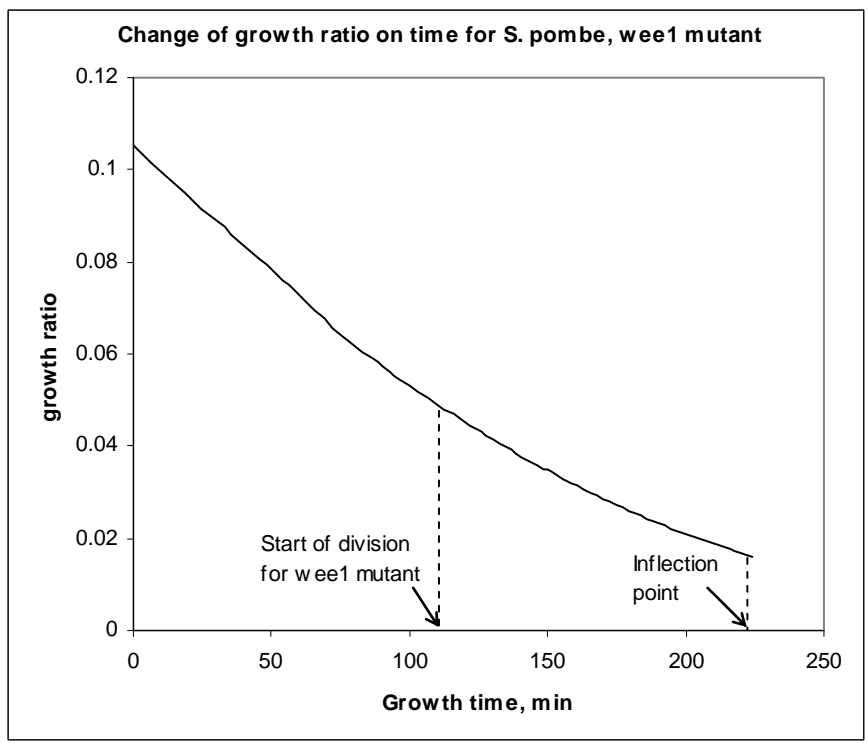

Fig. 14. Change of growth ratio for the $wee1\Delta$ mutant. The cell prematurely enters the division phase, before the inflection point is reached.

The appropriate growth curves that take into account influx required for protein and RNA synthesis are shown in Fig. 15. We presented nutritional scenarios from Table 3 with minimal (1) and maximum (4) amount of nutrients. The curves corresponding to scenarios 2 and 3 are located between these curves. The growth curve corresponding to minimal amount of nutrients has the smallest curvature, which is consistent with the results of experiments confirming that the growth curves of cells in nutritionally poor media are closer to linear than in the case of cells growing in a nutritionally rich medium. We can see a good correspondence between the experimental data and computed growth curve.

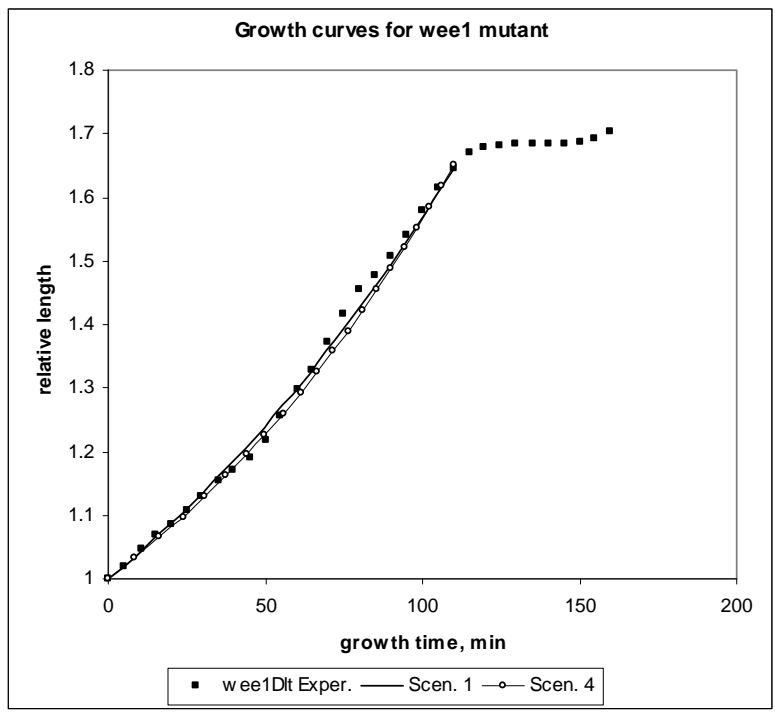

Fig. 15. Growth curves for the $wee1\Delta$ mutant without influx for DNA synthesis.

Let us take into account the influx required for DNA synthesis. DNA synthesis in case of small $wee1\Delta$ mutant has specifics. Because of the small size of the mutant cell, DNA synthesis does not start at the beginning of growth, like in the case of normal cells in a nutritionally normal environment. Rather, the cell first builds the aforementioned "DNA production plant" during a longer than usual G1 phase. Then, in the S phase, DNA synthesis begins. This is exactly what happens in the case of $wee1\Delta$ mutant.

Fig. 16 presents the case when influx for DNA synthesis is added. We can see that in this case the growth curve fits experimental data very well. The value of influx for DNA synthesis is about 7.5% of the total influx, on average, and we assume that the DNA synthesis begins at the point when mutant reaches the initial size of the wild type *fission yeast* that grows in normal nutritional conditions. (This is when the wild type *fission yeast* starts DNA synthesis immediately, or almost immediately.) The graph in Fig. 17 shows the appropriate influxes.

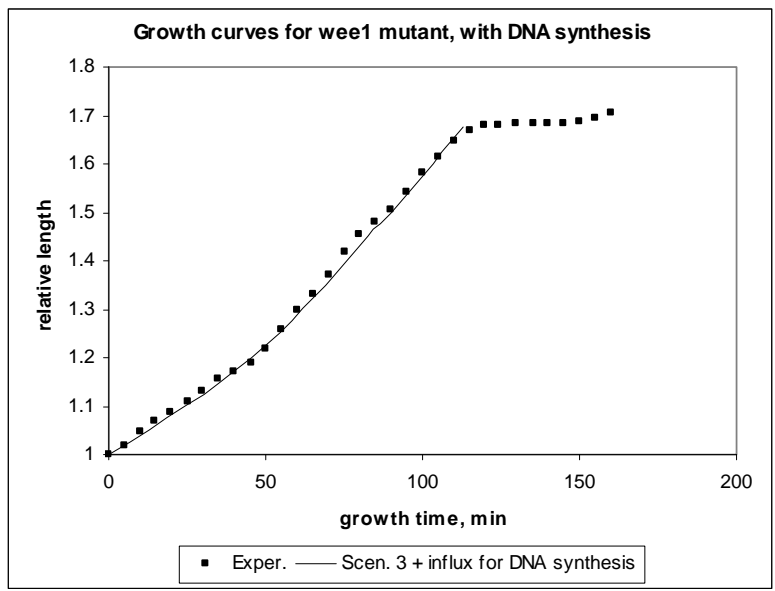

Fig. 16. Growth curve for the $wee1\Delta$ mutant with addition of influx for DNA synthesis.

So, overall, the growth curves which we computed based on very reasonable assumptions, such as the value of influx required for DNA synthesis and the delay of DNA synthesis because of the small size of the mutant cell, fit the experimental data very well. This one more time confirms the adequacy of the growth equation and validity of the discussed growth mechanism.

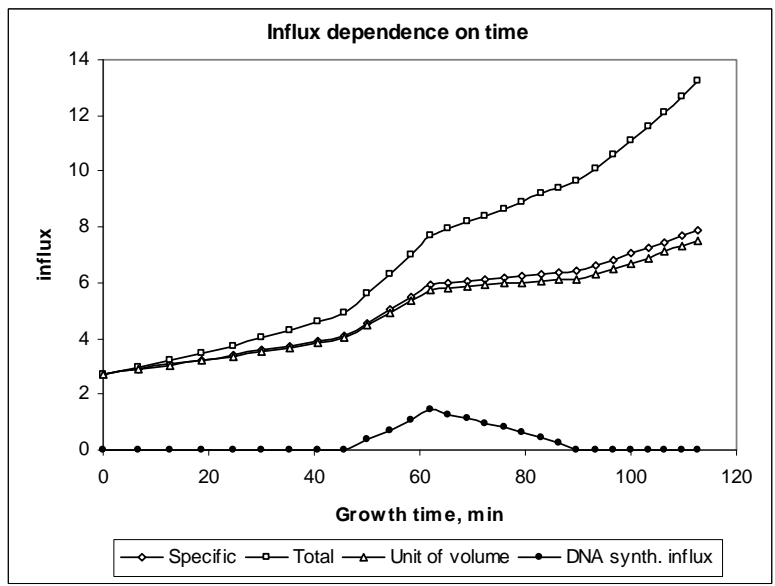

Fig. 17. Dependence of influx per unit of volume, specific and total influxes, and additional influx for DNA synthesis for $wee1\Delta$ mutant on time.

### 6.6. *The growth scaling mechanism revisited*

Previously, in subsection 4.7.2, we explained why cells of the same type can grow large and small based on the introduced growth scaling mechanism that follows from the general growth law. Also, we explained why actual growth can diverge from this scaling mechanism in case of extreme growth conditions. Since in this section we learned that there is a division mechanism of a different type when organisms switch to division at inflection point of the growth curve, we would like to make additional notes in this regard.

The growth curve in the area of inflection point is smooth. So, we should expect that there are certain divergences in sizes when division begins. However, these divergences have to be about the same, *in relative terms*, for small and large organisms of the same type, since they use the same division mechanism. On the other hand, organisms that use different division mechanisms, should exhibit different divergence in division sizes. In our case, we have two organisms that use the same division mechanisms, which are wild type *fission yeast* and diploid wild type *fission yeast*. So, we should expect that relative variance in sizes should be the same for both organisms. Indeed, according to Fig. 1B from Ref. 22, the variances in division sizes for them are 25% and 27%, which is very close, and, given measurement errors, within the error range. This fact is an additional proof in favor of the growth scaling mechanism that explains existence of small and large cells of the same type.

On the other hand, $wee1\Delta$ mutant uses different mechanism to switch to division, so that for it the variance in division size should be different. Given the fact that it is introduced by mutation, the variance should be greater compared to mechanism that is evolutionarily developed and optimized by Nature, because this mechanism is based on many interconnecting factors. Indeed, for the mutant, the variance in division sizes is 36%, which confirms our expectations. This value is far outside the range of possible errors, so that the estimate should be considered reliable.

Thus, we have solid experimental confirmation of two findings. First, the division mechanism is the same for small and large organisms of the same type. Second, we found an additional confirmation of workings of the growth scaling mechanism.

### 7. Growth of multicellular organisms

Above, we considered unicellular organisms. However, the general growth law is equally applicable to all living species and governs the growth of multicellular organisms, their systems and their organs. Let us take a look at the plum. The nutrients' influx enters closer to the center of the fruit and then it is distributed across the whole volume through a certain feeding system, based on capillary, osmotic and other physical, electromagnetic and biochemical effects. Each subsequent layer that surrounds the center obtains nutrients through the inner *surface* of this layer. Then, these nutrients are distributed in such a way that each unit of *volume* could receive the required amount of nutrients. So, essentially, we have the same surface – volume interaction, as in cells, although transformed into a different form. If we consider a blood system, then the inner *surface* of blood vessels supplies nutrients to the body's *volume*.

According to computations done for different geometrical forms, a sphere has the highest value of the growth ratio at the beginning. Accordingly, spherical fruits should grow faster. In this regard, we may anticipate two interesting things. First, fruits and vegetables that were evolutionarily developed in short growth seasons should have more round shape, because this is what allows them to grow faster. Second, nutrients have to be delivered as close to the center as possible, because this arrangement optimizes the length of nutrient distribution paths from many perspectives. So, generally the further North fruits and vegetables grow, the more round shape they should have because of the shorter vegetation season. Indeed, apples are more round than pears which grow in a warmer climate. Northern berries are round for the same reason. Although, unlike in apples, the stem is often attached to the berries surface, the arrangement, in fact, is the same, since many berries have sort of internal stem that first guides nutrients to the center. The further south we go, the more exotic shapes of fruits and vegetables are allowed, since the requirement of fast growth is not so urgent, and other factors enter the scene. Cucumber, a genially tropical vegetable, is long, because long shapes have slowly decreasing growth ratio and so such plants can grow for a long time. Elongated forms impose fewer limitations on the maximum possible size[1,3]. In fact, longer growth periods allow acquiring bigger mass, which is important for reproduction, especially in dry warm climates. However, the nutrient supply per unit of time can be limited, which is usually the case. When the rate of nutrient supply is limited, the only way of increasing mass is to increase the growth period, but then the shape has to be adjusted in a way that provides a slowly diminishing growth ratio. This is exactly what southern and tropical plants do, acquiring elongated or flattened forms. However, northern fruits and vegetables cannot do the same since their growth period is short. Optimizing the shape to fit a short growth period is priority number one for them. Thus, in case of multicellular fruits and vegetables, we also see evident workings of the general growth law. In fact, if we apply the growth equation to growth of spherical fruits, such as an apple, we will obtain growth curves similar to the ones computed in Ref. 3 for spherical unicellular organisms that obtain nutrients through their surface.

So, the growth equation can be readily applied to the growth of fruits and multicellular organisms in general. We just have to take into account that in this case we are dealing with several interacting systems that distribute resources. In fact, we have a *system* of growth equations, each describing a certain influx of resources for a certain organism's system and/or organ. For instance, in the human body, we have blood and respiratory systems which supply nutrients and oxygen and dispose of waste, as well as many other systems and organs, each of which uses a certain share of supplied resources and produces a certain amount of waste. During growth, each system divides the acquired resources between the maintenance needs and synthesis of new biomass. Systems are interdependent; in their development, they influence each other. Somebody cannot have a poorly developed cardio-vascular system and possess big and strong muscles. Because of these numerous feedback relationships, during normal growth, all systems and organs adjust to each other in terms of their consumption of resources and functional capacity; their development is interdependent and the functionality of each depends on other organs and systems. This is why the distribution of resources has to be so delicately balanced between the needs of different constituents of an organism.

Previously we found that in the particular case when nutrients are supplied through the surface, as in cells, the total influx is $K(V(X,t)) = k(X,t) \times S(X)$. The growth equation (1) can be rewritten for a single plum as follows. Instead of influx through the surface, we have to use the *total* influx of nutrients $K$ supplied through the fruit's stem. So, in general, when influx is not necessarily supplied through the surface, the growth equation should be written as follows.

$$p_c(X)dV(X,t) = K(X,t) \times \left(\frac{R_S}{R_V} - 1\right)dt \tag{31}$$

where $K(X,t)$ is the total influx. In other words, the growth of the organism is still defined by influx of nutrients and geometrical properties of the growing organism via the growth ratio. For the whole organism, the growth equation has to take into account the growth of all separate systems and organs, so that

$$p_c(X)dV(X,t) = \sum_{i=1}^{N} K_i(X,t) \times \left(\frac{R_{Si}}{R_{Vi}} - 1\right)dt \tag{32}$$

Here, index *i* relates to *i*-th system or organ. Some systems can obtain nutrients through the surface, such as lungs, which deliver oxygen. In others, the influx can come through the blood vessels or in some other way.

On the other hand, we can write a separate growth equation for each organ and system, so that we obtain a *system* of differential equations.

$$p_{ci}(X)dV_i(X_i,t) = K_i(X_i,t) \times \left(\frac{R_{Si}}{R_{Vi}} - 1\right)dt, \ i = \{1,..,N\} \tag{33}$$

From organs and systems we can go to the cell level, adding the growth equations for separate cells in order to have a better understanding of influx requirements and other specific features of certain growth processes. All of these equations interrelate to each other through the distribution of nutritional resources and flows of disposed materials. There are many degrees of freedom in this problem that can be beneficially used in order to find a correct and mathematically stable solution.

Besides, we can introduce additional constraints. For instance, the sum of all influxes is equal to the total influx of nutrients.

$$K(t) = \sum_{i=1}^{N} K_i(t) \tag{34}$$

We can have similar constraints for an organism's interdependent subsystems. Certainly, constraints may contain other information and considerations, including the organism's adaptation requirements and influence of the surrounding environment. Constraints also serve as connections and descriptions of relationships between different equations.

From the mathematical perspective, depending on what variables and functional dependencies are unknown, the system of equations (32), (33) that includes constraints, such as (34), can be underdetermined, overdetermined or well defined. Accordingly, depending on the situation, different mathematical methods can be used to find the solution.

Summarizing this section, we would like to turn the readers' attention to the fact that the discussed growth mechanism works at all levels of an organism: it shapes organelles, cells, organs, systems and the whole organism. In fact, the structure of organisms and their size and mutual synchronization of growth of different organs and systems is largely due to the workings of this general growth law during the evolution and developmental process, as well as genesis of an individual organism.

Overall, together with biochemical mechanisms, the general growth mechanism that we introduced governs growth and replication processes and defines the growth limits of cells, organs, systems and organisms.

## 8. Conclusion

In this article, we introduced and studied the general growth law. This law unites physical and biochemical mechanisms of growth into a coherent mechanism that universally governs the growth and replication of all living species. It turned out that the foundation of the general growth law is the existence of uniquely defined optimal distribution of nutritional resources between maintenance needs of existing organism and synthesis of new biomass. We found that the parameter, which we introduced as a growth ratio, is a mathematical representation of such optimal distribution, so that the growth ratio directly defines the fraction of nutrients that are used for biomass synthesis. The amount of synthesized biomass is an extremely important parameter, because it defines the composition of biochemical reactions. This composition always changes in such a way that it produces amount of biomass predefined by the growth ratio. Cumulatively, this ongoing change of composition of biochemical reactions is what forces an organism to proceed through the whole growth cycle and eventually replicate or switch to a quiescent state. Although the change of biomass toward the end of growth becomes smaller, the relative change of growth ratio, on the contrary, accelerates. These significant changes of the growth ratio, and the appropriately large changes in the relative amount of produced biomass, is what makes the replication of organisms such a definitive and well defined process with very rare failings. Such a division mechanism was developed in organisms that use the whole growth cycle predefined by the growth equation. In case of organisms that switch to division at inflection point of the growth curve, besides the growth ratio, additional factors help to trigger the division phase.

The growth ratio is not a magic number but quite an objective expression of an optimum balance between several competing and interacting processes. First, we have interaction between nutrient supply, either through the surface or in some other way, and the necessity to distribute these nutrients in order to support the functioning of biomass distributed in certain volume that has a certain geometrical form. This task had to be optimized in this competitive world, and it is, as we saw through the whole content of this paper. Second, the tasks of supporting existing biomass and creating new one is another pair of the inherently united and at the same time competing processes that can only exist together, as one inseparable unity. Indeed, interaction of these processes is also optimized and fine tuned by natural selection, and it could not be otherwise among the species competing for scarce resources. This is not an option, but the necessity and a matter of survival among other species which claim the same resources and striving to survive and maybe prosper. (Prospering what for? That's a question humans may ask (maybe some species too?), but not vegetables. It could be an interesting development, there are interesting and quite convincing answers to this question as well, but that study is beyond the scope of this paper.)

We convincingly proved the validity of the general growth law and its mathematical representation, the growth equation. For that, we explored four experimental growth dependencies: for *amoeba*, two experimental dependencies for wild type *fission yeast*, and *fission yeast* mutant $wee1\Delta$. In all cases, the growth curves computed on the basis of the growth equation corresponded to experimental data very well. The few reasonable assumptions that we used were well justified and supported by indirect information, evidence and experimental facts, as well as results obtained by other researchers. We discovered many other proofs of workings of the general growth law. In this regard, we present a credible scientific theory.

We found that there are two distinct types of growth scenarios supported by the appropriate biochemical mechanisms and geometrical characteristics of organisms. One type is presented by *amoeba*, which uses the whole growth cycle defined by the growth equation. Accelerating change of the growth ratio towards the end of growth, which affects the production of biomass through the change in composition of biochemical reactions, is the trigger that switches *amoeba* and similar organisms to the division phase. In the second type of growth, organisms realize the fastest possible growth scenario. Their growth proceeds according to the growth curve defined by the growth equation, until the inflection point. At this point, the organism switches to the division phase. However, the trigger to division is not so much the growth ratio, which still defines the composition of biochemical reactions throughout growth until the inflection point, but the deceleration of the growth rate coupled with biochemical mechanisms that work in such a way that they activate the accumulated mitotic material, including moving certain components, such as Pom1, toward the ends in order to reduce its concentration in the cortical nodes of the cell, which triggers chain of phosphorylation events starting mitosis.

In the case of the mutant, which switches to division earlier than a normal cell, because of the earlier activation of mitotic material, switching to division happens before the cell reaches the inflection point. All biochemical machinery that is required to reach a normal size is generally available in such mutants, but the part of machinery that is responsible for matching the transition to replication with reaching the inflection point on the growth curve is not working properly. This is why mutants can grow larger than normal cells or begin division when they are smaller.

The considered examples by far do not exhaust the possible applications of the growth equation and the growth mechanism. For instance, there are bacteria that live in water which was sealed inside stones. They managed to survive for centuries in an environment extremely poor in nutrients. Their growth cycle lasts for years because of the small amount of nutrients, which are minerals that gradually dissolve from the surrounding rock. In light of the growth mechanism, the existence of such bacteria and their extremely slow growth and replication cycle is very well founded. This is also information that might be of interest for gerontologists.

In fact, the so called "miracle of life" is overly exaggerated. Of course, it's an interesting natural phenomenon, but its core arrangement, meaning the action of the fundamental mechanisms that govern the existence and reproduction of living organisms, is not so complicated and certainly quite comprehensible. It has to be this way, because general laws have to embrace a very wide range of phenomena, and so they have to include minimum and only the most important fundamental parameters[26]. Even on a molecular level, once we understand the workings of the general laws, the molecular reactions become much more ordered and predictable.

**8.1.** *Generality of the physical growth mechanism*

Note how the growth equation, which mathematically is relatively simple in appearance, very adequately describes *real* growth processes. This is one of the properties of the right mathematical formulation of *general laws* of Nature. This is how general laws manifest themselves; they govern the core development and evolvement of natural phenomena. The elegance and simplicity of mathematical formulation of general laws is a consequence of their *generality*; the whole world we know is brought into existence and governed by these general laws interconnecting and transforming different forms of matter into each other. Because of their generality, they cannot be complicated; otherwise, they would not be general laws. The foundation of complexity is fragmentation, while generality is applied to the *whole,* as one monolithic piece, which generality has to be. The world may seem as a collection of random things to many of us. This is not so. Of course, randomness and determinism are two things that are inseparable. However, there is more determinism in this world than many think. We just don't know what forces shape it, what general laws govern the world's development. However, this lack of knowledge does not mean the absence of those forces and general laws. If not for these laws, the world would be chaos.

**Acknowledgements**

The author is thankful to Dr. Piotr H. Pawlowski from Institute of Biochemistry and Biophysics of Polish Academy of Sciences, Warszawa, and Dr. Peter Fantes for their invaluable support of this research.